\documentclass[aps,prx,reprint, amsmath, amssymb,superscriptaddress]{revtex4-2}

\usepackage{lmodern}
\usepackage[utf8]{inputenc}

\usepackage{bm}
\usepackage[retainorgcmds]{IEEEtrantools}
\usepackage{graphicx}
\usepackage{mathrsfs}
\usepackage{amsmath}
\usepackage{amssymb}
\usepackage{color}
\usepackage{amsfonts}
\usepackage{times,txfonts}
\usepackage{nicefrac}
\usepackage[colorlinks=true,linkcolor=blue,urlcolor=blue,citecolor=blue,pdfusetitle]{hyperref}
\usepackage{physics}
\usepackage{soul}
\usepackage{dsfont}
\usepackage{geometry}
\usepackage[dvipsnames]{xcolor}
\usepackage[authormarkup=none]{changes}

\usepackage{cancel}

\DeclareMathOperator*{\maxp}{max'}

\makeatletter
\newsavebox{\@brx}
\newcommand{\llangle}[1][]{\savebox{\@brx}{\(\m@th{#1\langle}\)}%
  \mathopen{\copy\@brx\kern-0.5\wd\@brx\usebox{\@brx}}}
\newcommand{\rrangle}[1][]{\savebox{\@brx}{\(\m@th{#1\rangle}\)}%
  \mathclose{\copy\@brx\kern-0.5\wd\@brx\usebox{\@brx}}}
\makeatother

\usepackage{tikz}

\begin{document}

\title{Emergence of Thermodynamics from Equilibration in Isolated  Quantum Systems}
\date{\today}
\author{Adalberto D. Varizi}
\affiliation{Instituto de Física, Universidade Federal Fluminense, Av. Litoranea s/n, Gragoatá 24210-346, Niterói, Rio de Janeiro, Brazil}
\author{Daniel Jonathan}
\affiliation{Instituto de Física, Universidade Federal Fluminense, Av. Litoranea s/n, Gragoatá 24210-346, Niterói, Rio de Janeiro, Brazil}
\author{Thiago R. de Oliveira}
\affiliation{Instituto de Física, Universidade Federal Fluminense, Av. Litoranea s/n, Gragoatá 24210-346, Niterói, Rio de Janeiro, Brazil}

\date{\today}

\begin{abstract}

Understanding how macroscopic thermodynamic behavior emerges from microscopic quantum dynamics remains an open problem.
While equilibration of quantum observables is well established, thermodynamics also relies on variables not directly associated with linear operators, but which are defined instead as functions of expectation values. 
Whether and how such derived quantities inherit equilibration properties is an open question.
Here, we establish that any continuously differentiable function of equilibrating expectation values also equilibrates.
We apply this result to a bipartite isolated system, showing that the entropy and conjugate variables of each subsystem --- defined through Jaynes’ maximum entropy principle --- equilibrate.
Moreover, with the assumption that their equilibrium properties depend solely on local conserved quantities, we show the dynamical maximization of the total entropy, enforcing equality of conjugate variables across subsystems.
These results provide a direct dynamical justification for entropy maximization and the emergence of thermodynamic equilibrium conditions, showing that fundamental principles of thermodynamics follow from the unitary evolution of quantum systems.
\end{abstract}
                              
\maketitle{}


\section{\label{sec:Intro}Introduction}


Recent decades have brought considerable progress in the understanding of the foundations of statistical mechanics and thermodynamic behavior from the underlying microscopic quantum dynamics of many-body systems.
For instance, typicality arguments show the expected values and variances of observables on the overwhelming majority of pure states of large systems closely match those predicted by the microcanonical ensemble~\cite{Lloyd2013,Gogolin2010,Gogolin2016}.
In particular, for sufficiently small subsystems, the reduced state is typically exponentially close to the canonical ensemble~\cite{Popescu2006,Goldstein2006}, a result that extends also to generalized notions of subsystems~\cite{Correia2024}.

Moreover, even for initial atypical nonequilibrium states, an isolated system with no extravagant degeneracies in its energy-gap spectrum exhibits effective equilibration.
More precisely, a series of well-known theorems~\cite{Reimann2008,Reimann2010,Reimann2012,Reimann2012a,Short2011,Short2012,Passos2025} show that the state will evolve in such a way that the expected value of any \emph{realistic observable} remains, at almost all times, close to stationary--- this is discussed in more detail in Sec.~\ref{sec:EquiOfRealObs}.
When restricting attention to observables supported on a small subsystem, the quantum dynamics will give rise to a reduced subsystem state that is for most of the time essentially indistinguishable from a time-independent state~\cite{Linden2009,Short2011,Short2012}.
Under the assumption of a weak coupling between this subsystem and its environment, it can be shown that this stationary state has the expected canonical form~\cite{Reimann2010}, thereby connecting dynamical equilibration with thermodynamic structure.

Concurrently, significant advances have been made in the understanding of how entropy (defined in a variety of different senses) increases or is produced along the quantum dynamics of systems~\cite{Esposito2010a,Polkovnikov2011b,Tasaki2016,Landi2020Rev,Strasberg2021,Varizi2024,Schindler2025}.
From an information-theoretic perspective, the production of entropy can be interpreted as arising from an observer’s limited ability to retrieve information that was initially accessible, given limited measurement capabilities.
Within this viewpoint, it was shown in~\cite{Varizi2024} that several seemingly distinct notions of entropy production can be unified through Jaynes’ maximum entropy principle~\cite{Jaynes1957,Wichmann1963}.

Despite these advances, the equilibration of expected values of observables alone does not suffice to fully characterize thermodynamic behavior.

First of all, a complete thermodynamic description also involves variables -- like entropy and temperature -- that are not themselves expectation values of observables, but can instead be defined as functions thereof.
This raises the question of whether the equilibration of expected values induces the equilibration of such derived thermodynamic variables. Perhaps surprisingly, to our knowledge, only relatively few results in this direction have been obtained~\cite{Ikeda2015,Meier2025,Schindler2025}, and only for certain entropy functions associated with coarse-grained measurements. 

In addition, even when all these variables equilibrate, there still remains the question of whether the resulting equilibrium values are thermal, in the sense that they agree with thermodynamic and statistical mechanics predictions.
In this regard, the most common justification for thermalization is the Eigenstate Thermalization Hypothesis~\cite{Jensen1985,Srednicki1994,Deutsch1991,Tasaki1998,Rigol2008,Gogolin2016}, although seemingly weaker assumptions may suffice~\cite{Sirker2014}.

In this work, we first give a simple argument extending the aforementioned equilibration theorems for expectation values~\cite{Reimann2008,Reimann2010,Reimann2012,Reimann2012a,Short2011,Short2012} to any differentiable function thereof.
Specifically, we provide bounds for their average squared fluctuations around equilibrium.
Moreover, by applying standard assumptions of textbook statistical mechanics to a bipartite isolated system, we show the emergent thermal behavior of an entropy function, subsystems' temperatures and other thermodynamic variables.

\begin{figure}[t]
    \centering
\includegraphics[width=1.\linewidth]{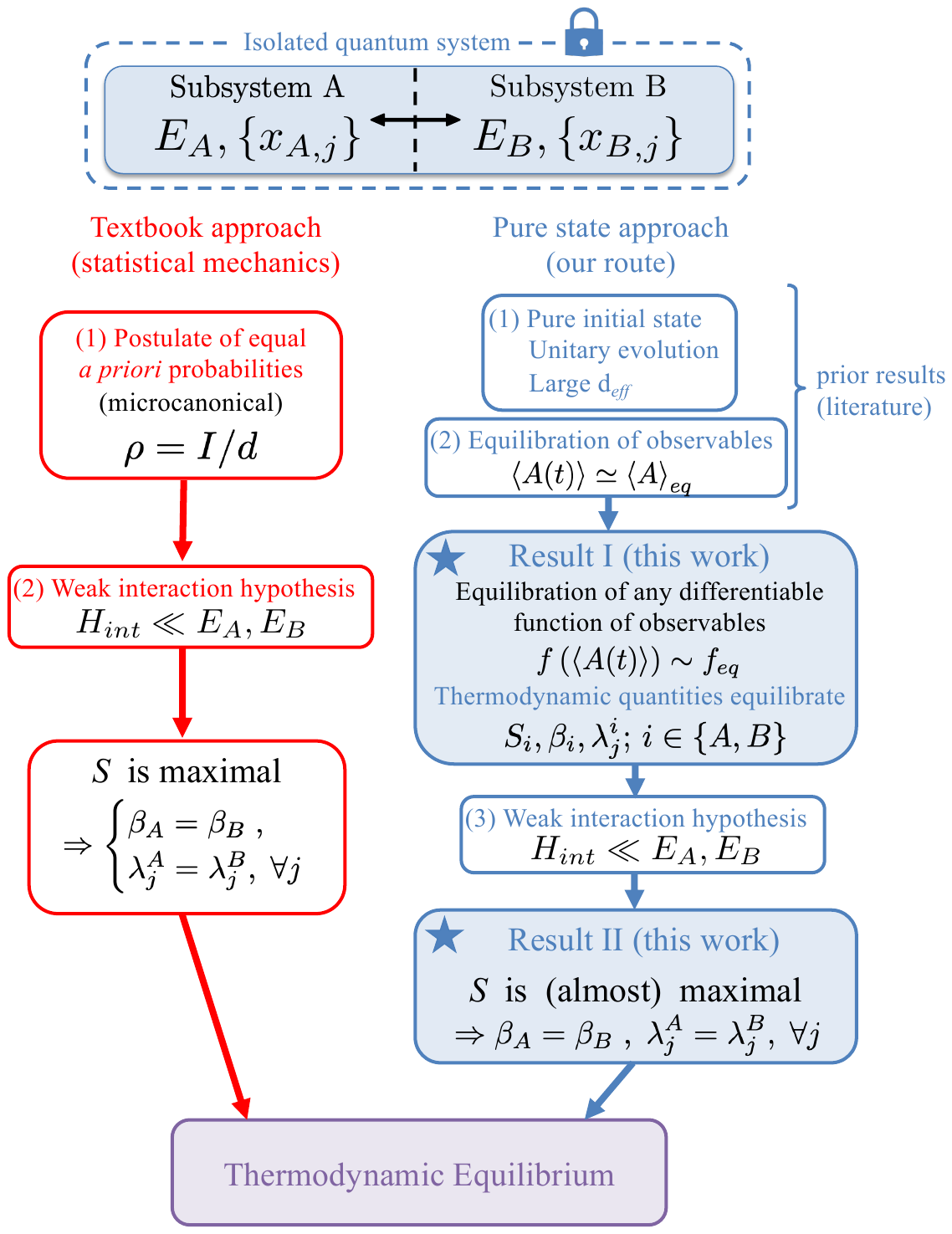}
    \caption{\label{fig:Summary}Schematic comparison between the textbook statistical-mechanics derivation of thermodynamic equilibrium and the approach developed in this work. The conventional route assumes the microcanonical postulate and weak interactions to infer entropy maximization and the equality of thermodynamic conjugate variables. Our route instead starts from the unitary evolution of an isolated quantum system in a pure state~\cite{Reimann2008,Short2011,Linden2009}. Using previous results on the equilibration of realistic observables~\cite{Reimann2008,Reimann2010,Reimann2012,Reimann2012a,Short2011,Short2012}, together with our two main results — (i) equilibration of any differentiable function of equilibrating observables and (ii) effective maximization of entropy under weak interactions — we recover thermodynamic equilibrium without assuming equal \emph{a priori} probabilities.}
\end{figure}

Explicitly, we consider two weakly coupled systems exchanging energy and other extensive quantities.
Using the maximum entropy principle, we define \emph{nonequilibrium} thermodynamic entropies and the corresponding entropy-conjugate variables, such as temperature, for each subsystem.
Assuming the equilibration of the exchanged quantities, we show the consequent equilibration of the entropies and the conjugate variables in each subsystem.
In addition, taking these exchanged quantities as the only thermodynamic relevant variables of the system, we show that the equilibration of the total entropy occurs effectively at its maximum, which in turn implies the equality of the equilibrium values of each conjugate variable across subsystems.
In Fig.~\ref{fig:Summary} we compare our approach to the usual route in statistical mechanics, where thermodynamic equilibrium is derived as a consequence of the equal probabilities postulate.

The paper is divided as follows.
In Sec.~\ref{sec:EquiOfRealObs} we revisit results on the equilibration on average of observables in closed quantum systems.
In Sec.~\ref{sec:EquiOfThermalVariables} we show how this result can be extended to functions of expected values of observables.
We then consider a closed quantum system composed of two weakly interacting subsystems, define entropy and entropy-conjugate variables for each subsystem and show that these quantities equilibrate.
In Sec.~\ref{sec:EquiToMaxEnt} we discuss the conditions for the equilibration of the total entropy of the system to occur at maximum and its relation to the notion of local and nonlocal conserved quantities.
In Sec.~\ref{sec:ToyModel} we use a one-dimensional spin model to illustrate our results.
We finally conclude in Sec.~\ref{sec:conc}.


\section{\label{sec:EquiOfRealObs}Equilibration of Experimental Observables}

Let us begin by reviewing the notion of equilibration of observables of isolated quantum systems \cite{Reimann2008,Reimann2010,Reimann2012,Short2011,Short2012,Reimann2012a}, and in particular its application to the class of so-called `experimentally realistic' observables, which are intended to model real measurement devices~\cite{Reimann2008,Reimann2010,Reimann2012,Reimann2012a}. 

Consider an isolated, finite quantum system with Hilbert space $\mathcal{H}$ and a time-independent Hamiltonian $H$ with finite spectrum:
\begin{equation}
    H = \sum_{n=1}^{d_E} E_nP_n,
\end{equation}
where $\{P_n\}$ is a complete set of orthogonal projectors and $d_E\le d=\mathrm{dim}(\mathcal{H})$ is the number of distinct energy eigenvalues $E_n$.
%
%
Let $\rho(t)=e^{-iHt}\rho(0)e^{iHt}$ denote the state of the isolated system at time $t$, given the initial contition $\rho(0)$,  and $A$ denote a generic observable of the system, whose expected value at $t$ reads $\tr\{A\rho(t)\}$. Furthermore, define
\begin{equation}
    \bar{\rho} = \sum_{n=1}^{d_E} P_n\,\rho(t)P_n = \sum_{n=1}^{d_E} P_n\,\rho(0)P_n,
\end{equation}
which can be viewed as the time-independent part of $\rho(t)$. Finally, let
\begin{equation}
    \overline{f(t)}^{\,\tau} = \frac{1}{\tau}\int_0^\tau \mathrm{d}t f(t),
\end{equation}
denote the time-average of an arbitrary function $f(t)$ in the interval $[0,\tau]$. 

A physically meaningful notion of equilibration  for observables can then be introduced as follows ~\cite{Reimann2008,Reimann2010,Reimann2012,Short2011,Short2012,Reimann2012a}: we say that $A$ equilibrates if there exists a time-independent state $\psi$ such that the differences
\begin{equation}\label{eq:ExpectDiff}
    \tr\{A\rho(t)\} - \tr\{A\psi\}
\end{equation}
between the actual expected values $\tr\{A\rho(t)\}$ and the stationary value $\tr\{A\psi\}$ are negligibly small for the overwhelming majority of times $t$ within a sufficiently large interval $[0,\,\tau]$, with only rare deviations.

In a closed quantum system, this is of course not generally true, for arbitrary choices of $A$. For example, if the initial state $\rho(0)$ is  the projector onto a superposition $\ket{E_1}+\ket{E_2}$ of two energy eigenstates with different energies, and $A = \rho(0)$ itself, then $\tr\{A\rho(t)\}$ oscillates forever between the values 0 and 1, never equilibrating.

Nevertheless, equilibration does indeed occur for a wide range of observables. 
One way to establish this is via the following bound, which holds for any observable $A$: for sufficiently large $\tau$, the time-average of the square of the difference in expectation values~\eqref{eq:ExpectDiff} satisfies~\footnote{The bound in Eq.~\eqref{eq:ReimannShortBound} combines results from~\cite{Reimann2012a,Short2012}. Indeed, one can show that $\overline{\tr\{A(\rho(t)-\bar{\rho})\}^2}^{\,\tau}\le||M||S$, where $S=\sum_{m\neq n}|\tr\{P_m\rho(t)P_n\,A\}|\le||A||^2\min\{1/d_\mathrm{eff},3\maxp_n p_n\}$ and $M$ is the matrix with entries $M_{mn,kl}=\overline{\exp[i(E_m-E_n-E_k+E_l)t]}^{\,\tau}$. $||M||\le\max_{kl}|M_{mn,kl}|$, with $M_{mn,kl}=1$ for degenerate gaps ($E_k-E_l=E_m-E_n$) and $|M_{mn,kl}|=4\sin^2[(E_m-E_n-E_k+E_l)\tau/2]/(E_m-E_n-E_k+E_l)^2\tau^2$ for nondegenerate gaps. Clearly, for sufficiently large $\tau$, the contributions from the nondegerenerate-gap terms can be made smaller than the contributions from the degenerate-gap cases. Hence $||M||\le2g$ for such sufficiently large times.}
\begin{equation}\label{eq:ReimannShortBound}
    \overline{\tr\{A(\rho(t)-\bar{\rho})\}^2}^{\,\tau} \le 2g ||A||^2\min\left\{\frac{1}{d_\mathrm{eff}},3\maxp_n  p_n\right\},
\end{equation}
where $g=\max_{m\neq n}\sum_{k\neq l}\delta_{E_k-E_l,\,E_m-E_n}$ is the maximal degeneracy of energy gaps~\footnote{That is, $g$ is the maximal number of equal energy differences among all possible pairs of distinct energy eigenvalues.}, $p_n = \tr\{P_n\,\rho(t)\}$ is the (time-independent) population of the energy subspace $E_n$; $\maxp_n p_n$ denotes the \emph{second} largest of these populations, and $d_\mathrm{eff}=1/\sum_{n=1}^{d_E}p_n^2\ge 1/\max_n p_n$ is the so-called effective dimension of the system, which measures how many energy levels contribute significantly to its state~\cite{Linden2009,Short2011,Short2012}.

The bound~\eqref{eq:ReimannShortBound} establishes that, provided that neither $g$ nor $||A||$ are exceptionally large, and that either $1/d_\mathrm{eff}$ or $\maxp_n p_n$ is sufficiently small, the expectation value $\tr\{A\rho(t)\}$ must remain close to the constant $\tr\{A\bar\rho\}$ over most times.

It turns out that, for realistic macroscopic systems and observables, these conditions are indeed typically satisfied. First of all, such systems are generically composed of many interacting parts, which suppresses the degeneracy of energy gaps, making $g$ comparatively small.  Notably, if all possible parts of the system interact with each other, such degeneracies are actually nonexistent~\cite{Linden2009}.


Furthermore, not all mathematically possible observables are in fact physically accessible in a macroscopic system. As argued in \cite{Reimann2008,Reimann2010,Reimann2012,Reimann2012a}, an `experimentally realistic' observable $A$, intended to model the output of any real measurement device, must have a finite range of possible outcomes, which implies a finite operator norm $||A||$, and also a finite resolution $\delta A$. 
In particular, relevant macroscopic quantities are often extensive, such that $||A||$ increases linearly with the system size $N$. At the same time, the spacing between their energy levels decreases exponentially with $N$~\cite[p. 28]{Reimann2010,LandauLifshitz}.



This leads to dramatic consequences for the distribution of energy populations~\cite{Reimann2008,Reimann2010,Reimann2012,Reimann2012a}: if the number of particles is of order $N=10^{23}$, and if one assumes a number of levels of order $10^{N}$ within each Joule of energy, then even an extremely small energy interval of size $\delta E =10^{-10^{22}}\,\mathrm{J}$ will contain $\ell=10^{9\times 10^{22}}$ energy eigenspaces~\cite{Reimann2010}.
The preparation, therefore, of a macroscopic system with even such a minuscule energy uncertainty as $\delta E$ will most likely lead to the occupation of a spectacularly massive number of energy eigenspaces~\cite{Reimann2008,Reimann2010,Reimann2012,Reimann2012a}.
For most macroscopic systems, then, the rough estimate~\cite{Reimann2010}
\begin{equation*}
    \max_n p_n \sim 10^{-N}
\end{equation*}
is quite reasonable~\footnote{Note that this exponential scaling for the decay of fluctuations with the system size is stronger than the $1/\sqrt{N}$ expected from textbook statistical mechanics.
This is due to entanglement and allows even small quantum system to equilibrate and thermalize~\cite{Correia2026}}.

Complementarily, general results on the geometry of high-dimensional Hilbert spaces show that states with small effective dimension $d_\mathrm{eff}$ are exponentially unlikely~\cite{Linden2009}.

Taken together, these properties ensure that, for experimentally realistic observables, the bound in eq. \eqref{eq:ReimannShortBound} is extremely tight. In other words, for most times, the expectation value $\tr\{A\rho(t)\}$ remains exponentially close to the equilibrium value $\tr\{A\bar\rho\}$.

This statement can be made more precise by considering the fraction of times for which deviations exceed the experimental resolution $\delta A$.
Let $\Delta t_{\delta A}$ denote the length of time within $[0, \tau]$ for which $|\tr\{A(\rho(t)-\bar{\rho})\}|\ge\delta A$.
Then, for sufficiently large $\tau$~\cite{Reimann2012,Reimann2012a},
\begin{equation}\label{eq:ReimanTimeBound}
    \frac{\Delta t_{\delta A}}{\tau} \le 2 g \frac{||A||^2}{\delta A^2}\min\left\{\frac{1}{d_\mathrm{eff}}, 3\maxp_n p_n\right\}.
\end{equation}

Since $||A||/\delta A$ scales polynomially with $N$ for any experimentally realistic observable, while $1/d_\mathrm{eff}$ and $\maxp_n p_n$ are typically exponentially small in macroscopic systems, this bound implies that deviations larger than the experimental resolution occur only during a negligible fraction of all times.
In this operational sense, the system equilibrates.

  

\section{\label{sec:EquiOfThermalVariables}Equilibration of Thermodynamic Conjugate Variables and Entropy}

The results reviewed in the previous section were geared toward understanding the equilibration of \emph{observables} represented by Hermitian operators.
However, when describing macroscopic systems and subsystems we are often interested in the equilibration of quantities which are \emph{not} of this kind, but are instead \emph{functions} of the state and/or of (realistic) observables.
Typical examples from thermodynamics include the (thermodynamic) entropy of the system, and also intensive quantities like temperature or pressure of subsystems.
In this section, we show how the results of section \ref{sec:EquiOfRealObs} can be leveraged to understand equilibration also in this sense.

More precisely: let $\{A_j\}_{j=1}^\ell$ be a set of experimentally realistic observables of an isolated quantum system, whose expected values we denote by $a_j(t)=\tr\{A_j\rho(t)\}$.
Assuming that all these observables equilibrate in the sense of the previous section, it  follows that any function
\[
    f:\mathbb{R}^\ell\to\mathbb{R},\quad (a_1,...,a_\ell)\mapsto f(a_1,...,a_\ell)
\]
that is continuously differentiable in the region
\[
    \mathcal{R} = \{r\,\mathbf{a}(t)+(1-r)\,\bar{\mathbf{a}}:\, t\in[0,\tau],\, r\in[0,1]\}
\]
where $\mathbf{a}(t)=(a_1(t),...,a_\ell(t))$ and $\bar{\mathbf{a}}=(\bar{a}_1,...,\bar{a}_\ell)$, with $\bar{a}_j=\tr\{A\bar\rho\}$, also equilibrates on average.

Indeed, since the expected values $\{a_j(t)\}$ evolve in time, let us denote by $f(t)=f(\{a_j(t)\})$ the value of $f$ at time $t$ and by $\bar{f}=f(\{\bar{a}_j\})$ the value of $f$ on the average vector.
The set $\mathcal{R}$ consists of all points lying on the line segments connecting the vector of expected values at time $t$, $\mathbf{a}(t)$, to the vector of equilibrium values $\bar{\mathbf{a}}$, for all $t\in[0,\tau]$.
Thus, if $f$ is continuously differentiable, by the Mean Value Theorem, for each time $t$, there exists a vector $\boldsymbol{\xi}(t)\in\mathcal{R}$ such that
\[
    f(t) - \bar{f} = \sum_{j=1}^\ell \frac{\partial f}{\partial a_j}\bigg|_{\boldsymbol{\xi}(t)} (a_j(t)-\bar{a}_j),
\]
with continuous and bounded first derivatives $\partial f(\boldsymbol{\xi}(t))/\partial a_j$.

Hence, let
\begin{equation}
    \mu_f = \max_{\displaystyle j} \max_{\textstyle t\in[0,\tau]} 
    \left\{\bigg|\frac{\partial f(\boldsymbol{\xi}(t))}{\partial a_j}\bigg|\right\}
\end{equation}
denote the maximal absolute value of any such derivatives $\partial f/\partial a_j$ among the points $\boldsymbol{\xi}(t)$.
Then
\[
    \begin{aligned}
        \left(f(t) - \bar{f}\right)^2 &= \left(\sum_{j=1}^\ell \frac{\partial f}{\partial a_j}\bigg|_{\boldsymbol{\xi}(t)} (a_j(t)-\bar{a}_j) \right)^2
        \\[0.2cm]
        &\le \left(\sum_{j=1}^\ell \bigg|\frac{\partial f}{\partial a_j}\bigg|_{\boldsymbol{\xi}(t)} |a_j(t)-\bar{a}_j| \right)^2
        \\[0.2cm]
        &\le \mu_f^2 \sum_{j,\,k=1}^\ell |a_j(t)-\bar{a}_j||a_k(t)-\bar{a}_k|,
    \end{aligned}
\]
where we used that $xy\le|x||y|$.

Now, since $\{A_j\}$ are experimentally realistic, we have
\begin{equation}\label{eq:epsilondef}
    \overline{\left( a_j(t)-\bar{a}_j \right)^2}^{\,\tau} \le ||A_j||^2\varepsilon,
\end{equation}
where $\varepsilon=2g\min\{1/d_\mathrm{eff}, 3\maxp_n p_n\}$ is the observable-independent factor in the equilibration bound~\eqref{eq:ReimannShortBound}.

Averaging over time and using the Cauchy-Schwarz inequality, we obtain
\begin{equation}\label{eq:mainI}
    \overline{\left(f(t)-\bar{f}\right)^2}^{\,\tau} \le \mu_f^2 \left(\sum_{j=1}^\ell ||A_j||\right)^2\varepsilon.
\end{equation}

\begin{figure}[t]
    \centering
    \includegraphics[width=1.\linewidth]{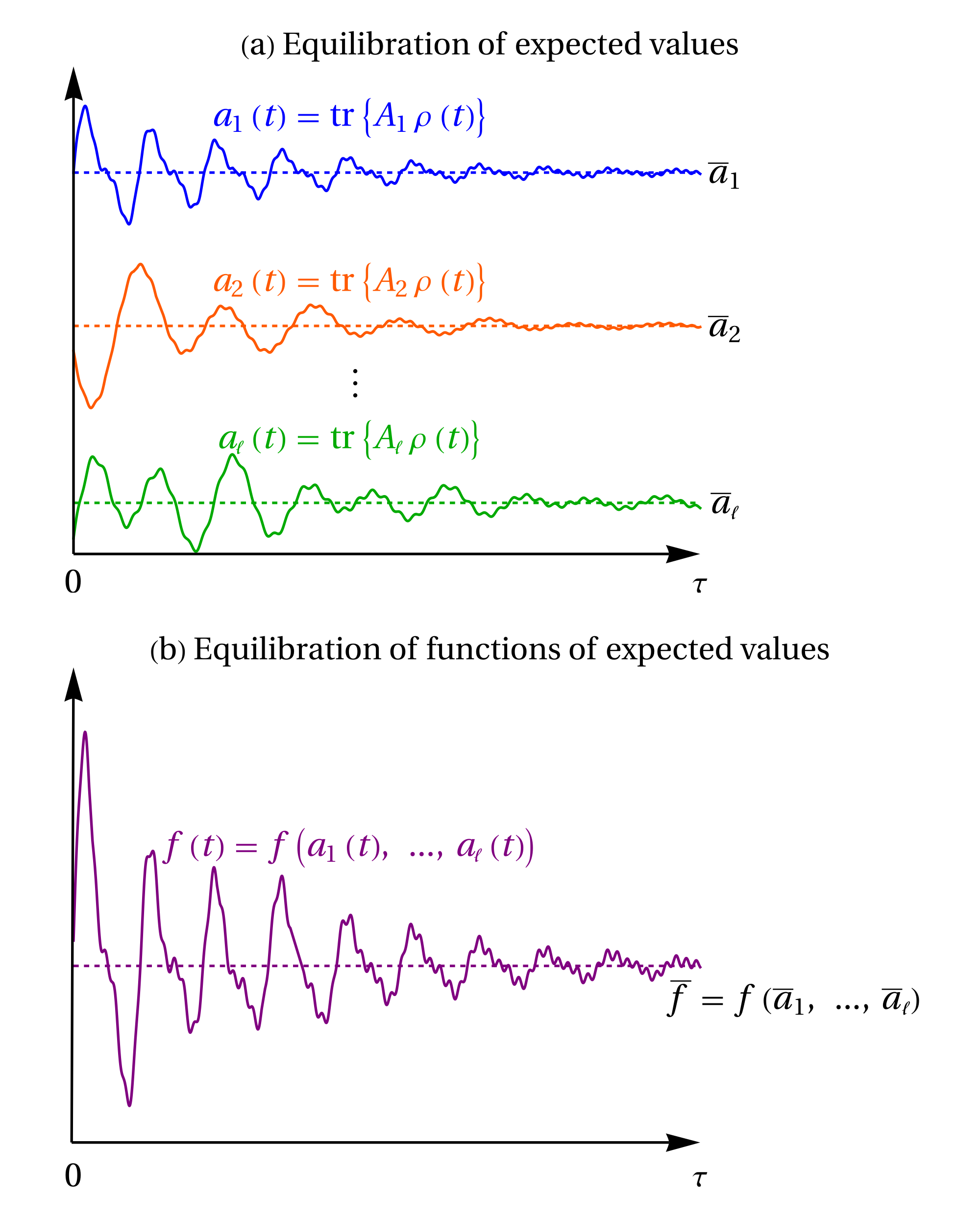}
    \caption{\label{fig:Equilibration}Equilibration of functions of equilibrating expected values. (a)  Illustration of experimentally realistic observables $A_j$ whose expected values $a_j(t)=\mathrm{tr}\{A_j\rho(t)\}$ equilibrate around stationary values $\bar a_j$. (b) A continuously differentiable function $f(t)=f(a_1(t),\ldots,a_\ell(t))$ constructed from the equilibrating expected values also equilibrates, around the stationary value $\bar f=f(\bar a_1,\ldots,\bar a_\ell)$.}
\end{figure}

Assuming any first derivative of $f$, and hence $\mu_f$, grows at most polynomially with the system size, this shows that $f(t)$ dynamically equilibrates to $\bar{f}$.
That is, $f(t)$ is, at almost all times, close to the value $\bar{f}$ associated with the equilibrium values of all observables $\{A_j\}$.
This is illustrated in Fig.~\ref{fig:Equilibration}.

Equation~\eqref{eq:mainI} is the first main result of this work.
It applies to any physical quantity that can be written as a function of the expected values of linear operators.

However, here we are mostly interested in the consequences of this result to the thermodynamic behavior of macroscopic systems. 
Generically, for a macroscopic \emph{isolated} system, it is known from experiments that a complete thermodynamic description and equilibrium characterization can be made in terms of a relatively small set of independent extensive quantities that are \emph{globally conserved}, but that can vary in time across the composing subsystems.
We call these quantities, the thermodynamically relevant quantities of the system. 
The archetypal quantity of this type is the energy itself: assuming the global isolated system is composed of many parts, the energies in each part may change in time while the global system energy remains fixed.

Once a system's set of relevant quantities is identified, other thermodynamic variables like entropy and temperature can be defined as functions of them.

Hence, let us consider the case where the isolated system is composed of two interacting subsystems $A$ and $B$, such that $\mathcal{H}=\mathcal{H}_A\otimes\mathcal{H}_B$ and --- see~\cite{note2again},
\begin{equation}\label{eq:ABHam}
    H = H_A + H_B + H_I,
\end{equation}
where $H_\alpha$, $\alpha=A,\,B$, denotes the time-independent Hamiltonian of subsystem $\alpha$ and $H_I$ describes their interaction.
In what follows, we assume the coupling between $A$ and $B$ is weak in the sense that, at any time, the total energy of the global isolated system is given by the sum of the energies of the subsystems. 
That is, let $E_\alpha(t)=\tr\{H_\alpha\,\rho(t)\}$ define the energy of subsystem $\alpha$ at time $t$, then the total (constant) energy $E=\tr\{H\,\rho(t)\}$ can be decomposed as~\footnote{Equivalently, we are assuming the energy associated with the interaction, $\tr\{H_I\,\rho(t)\}$, can always be ignored when compared to the energies of the subsystems $E_A$ and $E_B$.}
\begin{equation}\label{eq:weakcoupling}
    E\approx E_A(t) + E_B(t).
\end{equation}
This is of course the scenario that is generally assumed in statistical mechanics textbooks.
It is also usually valid in real macroscopic systems, often because subsystems only interact locally, over a boundary that is much smaller than the bulk.

Similarly, let $\{x_j\}_{j=1}^{d_T}$, with finite $d_T$, denote the other globally conserved quantities which, together with the energy $E$, allow a complete thermodynamic and characterization of the globally isolated system.
For instance, if this system is a spin chain, $x_1$ could represent the total magnetization along some given direction, and $x_1^A$ and $x_1^B$ this quantity's value for each respective subsystem.
If the system is an isolated gas partitioned by a rigid, permeable wall, $x_1$ could be the total number of particles, while $x_1^A$ and $x_1^B$ would be the fluctuating number of particles on each side of the wall.

As with the energy in~\eqref{eq:weakcoupling}, we assume that each $x_j$ can be decomposed as a sum of time-dependent values of the respective physical quantity in each subsystem:
\begin{equation}\label{eq:xdecomp}
    x_j = x_j^A(t) + x_j^B(t),
\end{equation}
and that each $x_j$ and $x_j^\alpha$ represent quantum mechanical expected values of observables, to which there are associated Hermitian operators $X_j$ and $X_j^\alpha$ such that --- again see~\cite{note2again} ---
\begin{IEEEeqnarray}{rCl}
    \label{eq:Xdecomp}
    X_j &=& X_j^A + X_j^B,
    \\[0.2cm]
    x_j &=& \tr\{X_j\rho(t)\},
    \\[0.2cm]
    x_j^\alpha(t) &=& \tr\{X_j^\alpha\,\rho(t)\},
\end{IEEEeqnarray}
with $X_j^\alpha$, $\alpha=A,\,B$, acting only on subsystem $\alpha$.

The bound~\eqref{eq:ReimannShortBound} implies quite generally that
\begin{IEEEeqnarray*}{rCl}
    \overline{\left( E_\alpha(t)-\bar{E}_\alpha \right)^2}^{\,\tau} \le ||H_\alpha||^2\,\varepsilon,
    \\[0.2cm]
    \overline{ \left( x_j^\alpha(t)-\bar{x}_j^\alpha \right)^2}^{\,\tau} \le ||X_j^\alpha||^2\,\varepsilon,
\end{IEEEeqnarray*}
with  $\bar{E}_\alpha = \tr\{H_\alpha\,\bar{\rho}\}$ the equilibrium energy, $\bar{x}_j^\alpha = \tr\{X_j^\alpha\,\bar{\rho}\}$ the equilibrium values of $\{x_j\}$-quantities of subsystem $\alpha$ and $\varepsilon$ defined as in~\eqref{eq:epsilondef}.
Assuming $\{H_\alpha, \{X_j^\alpha\}\}$, $\alpha=A,\,B$ are all experimentally realistic observables, we can expect these bounds to be exponentially small in $N$, and that these quantities equilibrate in the sense discussed in Section~\ref{sec:EquiOfRealObs}. 

The thermodynamic analysis of any physical system requires the introduction of an entropy functional for it.
A particularly useful approach in this regard is provided by Jaynes' maximum entropy principle~\cite{Varizi2024}.
This principle states that, given a limited set of known expectation values, the least biased description of a system is provided by the ensemble that maximizes the von Neumann entropy $S(\rho)=-\tr\{\rho \ln \rho\}$ subject to these constraints~\cite{Jaynes1957,Wichmann1963,Ingarden1997}.

Based on the recognition of $\{E,\{x_j\}\}$ as the thermodynamically relevant variables of the system, we define the time-dependent thermodynamic entropy of subsystem $\alpha$ as the maximum von Neumann entropy constrained by the expected values $\{E_\alpha(t),\{x_j^\alpha(t)\}\}$.  
As a consequence, we have --- see Appendix~\ref{AppI},
\begin{IEEEeqnarray}{rCl}
    \label{eq:entA}
    S_A(t) = \ln Z_A(t) + \beta_A(t) E_A(t) + \sum_{j=1}^{d_T}\lambda_j^A(t) x_j^A(t),
    \\[0.2cm]
    \label{eq:entB}
    S_B(t) = \ln Z_B(t) + \beta_B(t) E_B(t) + \sum_{j=1}^{d_T}\lambda_j^B(t) x_j^B(t),
\end{IEEEeqnarray}
where $\beta_\alpha$ and $\{\lambda_j^\alpha\}$ are given implicitly by
\begin{IEEEeqnarray}{rCl}
\label{eq:noneqTemp}
    E_\alpha = -\frac{\partial}{\partial \beta_\alpha}\ln Z_\alpha \left(\beta_\alpha,\{\lambda_j^\alpha\} \right),
    \\[0.2cm]
\label{eq:noneqXjs}
    x_j^\alpha = -\frac{\partial}{\partial \lambda_j^\alpha} \ln Z_\alpha \left(\beta_\alpha, \{\lambda_j^\alpha\} \right),
\end{IEEEeqnarray}
and
\begin{equation}
 Z_\alpha = \tr\left\{\exp(-\beta_\alpha H_\alpha - \sum_{j=1}^{d_T}\lambda_j^\alpha X_j^\alpha) \right\}.      
\end{equation}
In~\eqref{eq:entA} and~\eqref{eq:entB} we have made explicit the time dependence of $Z_\alpha$, $\beta_\alpha$ and $\{\lambda_j^\alpha\}$, which is inherited from that of $E_\alpha$ and $\{x_j^\alpha\}$.
It is also worth noting that $Z_\alpha, ~\beta_\alpha $ and $\{\lambda_j^\alpha\}$ should remain finite in any real physical system.
As discussed in more detail in Appendix~\ref{AppI}, the combination of the extensive and experimentally realistic characters of $\{H_\alpha,\{X_j^\alpha\}\}$ ensures that this is indeed the case.

The reasoning leading to the previous definition of the subsystem' entropy closely parallels the standard construction of the canonical ensemble in equilibrium statistical mechanics.
There, a Lagrange multiplier $\beta$ is introduced to enforce a specific mean energy $E$, but its identification with the derivative of entropy with respect to energy, $\beta=\partial S/\partial E$, imposes its recognition as the inverse of the system temperature.

Analogously, here the multipliers $\beta_\alpha$ and $\lambda_j^\alpha$ arise in association with the quantities $E_\alpha$ and $x_j^\alpha$, respectively, and moreover,
\begin{IEEEeqnarray}{rCl}
 \label{eq:betadef}
    \frac{\partial S_\alpha}{\partial E_\alpha} = \beta_\alpha,
    \\[0.2cm]
 \label{eq:lambdef}
    \frac{\partial S_\alpha}{\partial x_j^\alpha} = \lambda_j^\alpha.
\end{IEEEeqnarray}

We therefore take $\beta_\alpha(t)$ as the definition of the inverse temperature of subsystem $\alpha$ at time $t$, and give similar thermodynamic interpretations to the other entropy conjugate variables $\lambda_j^\alpha(t)$.
That is, if $x_j^\alpha(t)$ represents, for instance, the number of particles in subsystem $\alpha$, we assume $\lambda_j^\alpha(t)$ is related to the chemical potential of this subsystem.
We emphasize that all these quantities remain well-defined outside of equilibrium.

In fact, the notion of nonequilibrium temperature defined by~\eqref{eq:noneqTemp}, already appears as far back as in~\cite{Muschik1977,Muschik1977a}, where it was interpreted as the temperature of the bath that, when coupled to the system, results in no net heat exchange between them.
The same definition was also used in a derivation of first and second laws of thermodynamics for open and closed nonequilibrium quantum systems in~\cite{Strasberg2021}, where the authors further discuss its historical use in the literature. 

Finally, again following usual textbook assumptions, we define the total (thermodynamic) entropy of the isolated system as the sum of the entropies of its parts:
\begin{equation}\label{eq:bipentropy}
    S(t) = S_A(t)+S_B(t).
\end{equation} 

We may now apply the bound~\eqref{eq:mainI} to the entropy $S_\alpha(t)$ and the conjugate variables $\beta_\alpha$ and $\{\lambda_j^\alpha\}$.
In particular, for the inverse temperature $\beta_\alpha(t)$ we have
\begin{equation}\label{eq:betaBDeq}
    \overline{\left(\beta_\alpha(t) - \bar{\beta}_\alpha\right)^2}^{\,\tau} \le \mu_\alpha^2 \left(||H_\alpha|| + \sum_{j=1}^{d_T}||X_j^\alpha||\right)^2\varepsilon,
\end{equation}
where $\bar\beta_\alpha$ is the inverse temperature associated with the $\alpha$-subsystem equilibrium energy $\bar E_\alpha$ and other expected values $\{\bar{x}_j^\alpha\}$.
The coefficient $\mu_\alpha$ is given by
\[
    \mu_\alpha =  \max_{\displaystyle j} \max_{  \displaystyle t\in[0,\,\tau]}  \left| \left[C^{-1}_\alpha \left( \boldsymbol{\xi}_t \right) \right]_{1(1+j)} \right|,
\]
where $\left\{ \boldsymbol{\xi}_t \right\}$ are the mean-value points connecting the differences $\beta_\alpha(t) - \bar{\beta}_\alpha$ to the differences $E_\alpha(t) - \bar{E}_\alpha$ and $\{x_j^\alpha(t) - \bar x_j^\alpha\}$.
Moreover --- as we show in Appendix~\ref{AppCo} --- $C_\alpha^{-1}$ is the inverse of the matrix $C_\alpha$ with elements
\[
    \left[ C_\alpha \right]_{ij} = - \frac{\partial^2 \ln Z_\alpha}{\partial \lambda_i^\alpha \partial \lambda_j^\alpha},
\]
with $\lambda_{0}^\alpha \equiv \beta_\alpha$.
Noticeably, for extensive $H_\alpha$ and $\{X_j^\alpha\}$, $\left [C_\alpha \right]_{ij}$ is also typically extensive, which means $\mu_\alpha$ scales inversely with the system size and, therefore, that the right-side of~\eqref{eq:betaBDeq} has the same exponentially small scaling as $\varepsilon$.

Equation~\eqref{eq:betaBDeq} thus shows that the inverse temperature $\beta_\alpha(t)$ dynamically equilibrates and is close to $\bar\beta_\alpha$ at almost all times, as a result of the equilibration of all thermodynamic variables $\{E_\alpha,\{x_j^\alpha\}\}$ of the subsystem.

An analogous bound ensures the equilibration of all other conjugate variables $\lambda_j^\alpha$:
\begin{equation}\label{eq:lambBDeq}
    \overline{\left(\lambda_j^\alpha(t) - \bar{\lambda}_j^\alpha\right)^2}^{\,\tau} \le \mu_{\alpha j}^2 \left(||H_\alpha|| + \sum_{j=1}^{d_T}||X_j^\alpha||\right)^2\varepsilon,
\end{equation}
where -- again see Appendix~\ref{AppCo} --
\[
    \mu_{\alpha j} = \max_{\displaystyle l} \max_{\displaystyle t\in[0,\,\tau]}  \left| \left[C^{-1}_\alpha \left( \boldsymbol{\xi}_t^j \right) \right]_{j(1+l)} \right|,
\]
with $\left\{ \boldsymbol{\xi}_t^j \right\}$ the set of mean-value points connecting $\lambda_j^\alpha(t)-\bar \lambda_j^\alpha$ to $E_\alpha(t) - \bar E_\alpha$ and $\{x_j^\alpha(t) - \bar x_j^\alpha\}$.

In the case of the entropy $S_\alpha$, let $\bar{S}_\alpha = \ln \bar Z_\alpha + \bar\beta_\alpha \bar E_\alpha + \sum_{j=1}^{d_T}\bar\lambda_j^\alpha \bar x_j^\alpha$ denote the value corresponding to the equilibrium energy $\bar E_\alpha$ and with $\{\bar x_j^\alpha\}$.
Moreover let $\left\{\boldsymbol{\xi}_t^{\,S_\alpha} \right\}$ denote the set of mean-value points lying on the line segments between $S_\alpha(t)$ and $\bar{S}_\alpha$.
Recalling the first derivatives of $S_\alpha$ are the conjugate variables $\beta_\alpha$ and $\lambda_j^\alpha$ and denoting by
\[
 \lambda_{\alpha} = \max_{\displaystyle t\in[0,\tau]} \max\left\{|\beta_\alpha(\boldsymbol{\xi}_t^{\, S_\alpha})|,\,|\lambda_1^\alpha(\boldsymbol{\xi}_t^{\, S_\alpha})|,\,\dots,\,|\lambda_{d_T}^\alpha(\boldsymbol{\xi}_t^{\, S_\alpha})|\right\},
\]
the maximum, at any time $t\in[0,\,\tau]$, of the maximum absolute value of any of these derivatives at the mean-value points $\left\{ \boldsymbol{\xi}_t^{\,S_\alpha} \right\}$, we obtain
\begin{equation}~\label{eq:entABBD}
    \overline{\left(S_\alpha(t)-\bar{S}_\alpha\right)^2}^{\,\tau} \le \lambda_{\alpha}^2 \left(||H_\alpha|| + \sum_{j=1}^{d_T}||X_j^\alpha||\right)^2\varepsilon.
\end{equation}

Since the entropies of both subsystems equilibrate, so does the total entropy~\eqref{eq:bipentropy} and we have,
\begin{equation}\label{eq:entBD}
    \overline{\left(S(t)-\bar{S}\right)^2}^{\,\tau} \le \sum_{\alpha= A, B} \lambda_\alpha^2 \left(||H_\alpha|| + \sum_{j=1}^{d_T}||X_j^\alpha||\right)^2\varepsilon.
\end{equation}
That is, the entropies of each subsystem $S_\alpha(t)$, as well as the total entropy $S(t)$, equilibrate and appear stationary at almost all times.
In Appendix~\ref{AppAltBounds} we also give bounds of the same order as~\eqref{eq:entABBD} and~\eqref{eq:entBD} based on the concavity of $S_\alpha$.
These concavity-based bounds depend directly on the evolution of the Lagrange multipliers $\{\beta_\alpha(t),\, \{\lambda_j^\alpha(t)\} \}$ instead of on their values at $\{\boldsymbol{\xi}_t^{S_\alpha}\}$.
This would make their estimation easier.


\section{\label{sec:EquiToMaxEnt}Equilibration at Maximum Entropy}

Given generic nonequilibrium initial values for $\{E_A,\{x_j^A\}\}$ and $\{E_B,\{x_j^B\}\}$, the weakly coupled subsystems $A$ and $B$ will exchange energy and the other quantities $\{x_j\}$ until the emergence of an effective equilibrium.
The fundamental problem of thermodynamics is that of determining the conditions of this equilibrium.

Thus far we have shown that the equilibration of the set of variables $\{E_\alpha,\{x_j^\alpha\}\}$, $\alpha=A,\,B$, leads to the equilibration of the total entropy $S$.
In principle, however, this does not necessarily imply, for instance, the equality of the subsystems' equilibrium temperatures, $\bar\beta_A=\bar\beta_B$, or that the equilibrium value of $S$, i.e. $\bar{S}$, is maximum.

In the previous section we assumed the coupling between $A$ and $B$ is weak and that the globally conserved quantities $E$ and $\{x_j\}_{j=1}^{d_T}$ completely characterize the thermodynamics and equilibrium of the system.
Under these assumptions, we show here that the equilibration of the entropy $S(t)$ in~\eqref{eq:bipentropy} does occur effectively at maximum, and that, in turn, this leads to the equality of entropy conjugate variables on both subsystems.
In other words, we recover the expected thermodynamic results for equilibration.
This is the second main result of this work.

We first notice that, by construction, the maximization of $S$ truly enforces the equality of the subsystems' conjugate variables.
Indeed, under the assumptions that $E_A + E_B \approx E$ and $x_j^A + x_j^B = x_j$, for $j\in\{1,...,d_T\}$, i.e., that these quantities are all globally conserved, $S$ can be regarded as a function solely of $\{E_A,\{x_j^A\}\}$ (or, equivalently of $\{E_B,\{x_j^B\}\}$).
This allows us to obtain the textbook-like result,
\begin{equation}\label{eq:textbookEquilibEq}
 \begin{aligned}
    dS(E_A,\{x_j^A\}) &= dS_A(E_A,\{x_j^A\}) + dS_B(E_B(E_A),\{x_j^B(x_j^A)\})
    \\[0.2cm]
    & = \left( \beta_A -\beta_B \right)dE_A + \sum_{j=1}^{d_T} \left( \lambda_j^A - \lambda_j^B \right)dx_j^A,
 \end{aligned}
\end{equation}
where we used Eqs.~\eqref{eq:betadef} and~\eqref{eq:lambdef} and the fact that $dE_B\approx-dE_A$ and similarly $dx_j^B=-dx_j^A$.
Hence when $S$ is at maximum, this means that
\begin{equation} \label{eq:equiconds}
 \beta_A = \beta_B, \quad \lambda_j^A = \lambda_j^B,   
\end{equation}
for all $j\in\{1,...,d_T\}$.

Thus, let us denote by $S_\mathrm{max}$ the maximum possible value of $S$, which enforces Eqs.~\eqref{eq:equiconds}.
If we can show that the equilibrium value $\bar S$ is sufficiently close to $S_\mathrm{max}$, we consequently show that $S$ truly behaves as thermodynamically expected, and that the equalities~\eqref{eq:equiconds} hold at almost all times, while the system is effectively at equilibrium.
Our strategy to show that this is indeed the case will be the following:
We will first show that $S_\mathrm{max}$ is asymptotically indistinguishable from a value $S_\omega$.
Next we will establish what are the conditions for $\bar S$ to become asymptotically indistinguishable from $S_\omega$.
Since we will have $\bar S \approx S_\omega$ and $S_\omega \approx S_\mathrm{max}$, when these conditions are satisfied, $\bar S$ will also be asymptotically indistinguishable from $S_\mathrm{max}$.
Table~\ref{tab:TableOfEnts} highlights the differences between, and meanings of, $\bar S$, $S_\mathrm{max}$ and $S_\omega$.
\begin{table*}[t]
\centering
\renewcommand{\arraystretch}{1.2}

\begin{tabular}{c||p{4cm}||p{5.5cm}||p{4cm}}

Entropy &
\multicolumn{1}{c||}{Constraints} & \multicolumn{1}{c||}{Associated Maximum Entropy Ensemble} & \multicolumn{1}{c}{Meaning} \\
\hline

\begin{minipage}[c]{1cm}\centering$\bar S$\end{minipage} & \centering $\{\bar E_A,\{\bar x_j^A\}\}$ and
$\{\bar E_B,\,\{\bar x_j^B\}\}$ & \begin{minipage}{5.5cm}\centering $\bar\sigma\sim e^{-\bar\beta_A H_A-\sum_j \bar\lambda_j^\alpha X_j^A} \otimes e^{-\bar\beta_B H_B-\sum_j \bar\lambda_j^\alpha X_j^B} $ \end{minipage} & Equilibrium value of $S$~\eqref{eq:bipentropy}.
\\
\hline

\begin{minipage}[c]{1cm}\centering $S_{\mathrm{max}}$\end{minipage} & \centering $\beta_A=\beta_B$ and $\lambda_j^A=\lambda_j^B$ &
\begin{minipage}{5.5cm} \centering $\sigma_{\mathrm{max}} \sim
e^{-\beta_{\mathrm{max}}(H_A+H_B) -\sum_j\lambda_j^{\mathrm{max}} (X_j^A+X_j^B)}$ \end{minipage} & Max. allowed value of $S$~\eqref{eq:bipentropy}.
\\
\hline

\begin{minipage}[c]{1cm}\centering $S_\omega$ \end{minipage} & \centering $E$ and $\{x_j\}$. &
\begin{minipage}{5.5cm} \centering $\omega \sim e^{-\beta H-\sum_j\lambda_j X_j}$ \end{minipage} & Asymptotic value of both $\bar S$ and $S_\mathrm{max}$ for large $N$.
\\
\hline

\end{tabular}

\caption{\label{tab:TableOfEnts}Comparison between the entropies $\bar S$, $S_\mathrm{max}$ and $S_\omega$.
In the constraints column, $\{\bar E_\alpha,\{\bar x_j^\alpha\}\}$, $\alpha = A,\, B$ are the equilibrium values of the thermodynamically relevant quantities in each subsystem.
$S_\mathrm{max}$ is constrained by $\{E_A,\{x_j^A\}\}$ and $\{E_B\approx E-E_A,\{x_j^B=x_j-x_j^A\}\}$ that result in the conditions for maximum~\eqref{eq:equiconds}. Finally, $E$ and $\{x_j\}$ are the global, thermodynamically relevant, conserved quantities.}
\end{table*}

\begin{figure}[t]
    \centering
    \includegraphics[width=1\linewidth]{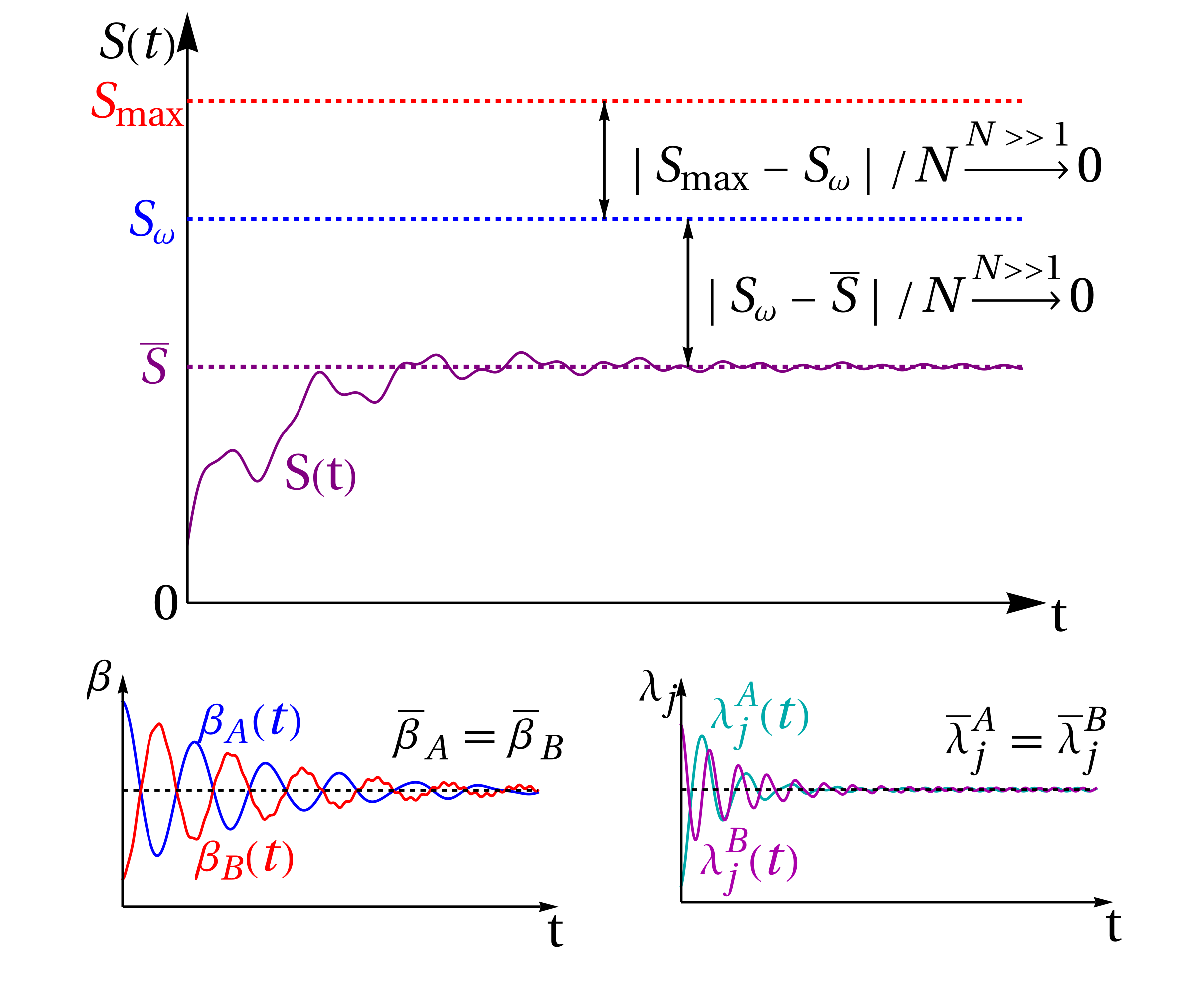}
    \caption{\label{fig:figIV} Schematic illustration of the emergence of thermodynamic equilibrium. The total entropy $S(t)$ equilibrates around its long-time average $\bar S$, here shown for a finite $N$. For sufficiently large systems, however, the entropy density $\bar S/N$ becomes asymptotically indistinguishable from the maximum attainable entropy density $S_\mathrm{max}/N$ and from the reference entropy density $S_\omega/N$. Consequently, $S(t)$ is effectively maximal at almost all times after equilibration. The entropy conjugate variables simultaneously equilibrate, leading to the equalities $\bar \beta_A = \bar \beta_B$ and $\bar \lambda_j^A = \bar \lambda_j^B$, which characterize thermodynamic equilibrium between subsystems $A$ and $B$.}
\end{figure}

We first recall that, for each value of $S$, there is and associated maximum entropy ensemble $\sigma$ whose von Neumann entropy reproduces this value: $S=S(\sigma)$ --- see Appendix~\ref{AppI}.
Specifically, this ensemble has the product form
\[
 \sigma = \frac{e^{-\beta_AH_A -\sum_j \lambda_j^AX_j^A } }{Z_A}\otimes\frac{e^{-\beta_B H_B -\sum_j \lambda_j^B X_j^B } }{Z_B},
\]
and satisfies $E_\alpha=\tr\{H_\alpha\sigma\}$ and $x_j^\alpha=\tr\{X_j^\alpha\sigma\}$.
Let $\sigma_\mathrm{max}$ be the ensemble associated with the maximum value of $S$: $S_\mathrm{max}=S(\sigma_\mathrm{max})$.
Under the weak coupling condition, $||H_I||/N\to0$ for sufficiently large $N$, we show in Appendix~\ref{AppIII} that 
\[
 \frac{1}{N}|S_\omega - S_\mathrm{max}| \le 2 \frac{||H_I||}{N} \max\{|\beta|,|\beta_\mathrm{max}|\} \to 0,
\]
where
\[
 \begin{aligned}
 &\omega = \frac{1}{Z}\exp(-\beta H - \sum_{j=1}^{d_T} \lambda_j X_j),
 \end{aligned}
\]
with $Z=\tr\{\exp(-\beta H - \sum_{j=1}^{d_T}\lambda_j X_j)\}$, and
\begin{equation}\label{eq:maxentprime}
    S_\omega \equiv S(\omega) = \ln Z + \beta E + \sum_{j=1}^{d_T}\lambda_j x_j
\end{equation}
is the maximum von Neumann entropy constrained by the global conserved quantities $E$ and $\{x_j\}$.
To emphasize, $S_\omega$ differs from $S$ by the fact that the latter is constrained not by the global, but by the subsystems' values $\{E_\alpha,\{x_j^\alpha\}\}$. 
Therefore, at finite $N$, $S_\omega$ may be smaller or bigger than $S_{\max}$, depending on the interaction $H_I$ (see Figs.~\ref{fig:figIV} and~\ref{fig:EntrGrid}). However, as $N\rightarrow \infty$ their respective densities converge.

Simultaneously, also under the weak coupling assumption, we show in Appendix~\ref{AppIII} that
\begin{equation}\label{eq:entdiff}
 \frac{|S_\omega - \bar S|}{N} \le \left|\bar \beta_A \!-\! \bar \beta_B\right| \frac{\left|\bar E_A \!-\! E_A^\omega\right|}{N} +\! \sum_{j=1}^{d_T} \left|\bar \lambda_j^A \!-\! \bar \lambda_j^B\right| \frac{\left|\bar x_j^A \!-\! {x_j^\omega}^A \right|}{N},
\end{equation}
where $E_\alpha^\omega=\tr\{H_\alpha\omega\}$ and ${x_j^\omega}^\alpha=\tr\{X_j^\alpha\omega\}$.

That is, in the limit of large $N$, the difference between the equilibrium value of $S(t)$, given by $\bar S$, and $S_\omega$ is bounded by the differences
\begin{IEEEeqnarray}{rCl}
    \label{eq:MaxEntCond1}
    \Delta E_\alpha^\omega = \left| \bar{E}_\alpha - E_\alpha^\omega \right| &=& \left| \tr\{H_\alpha(\bar\rho - \omega)\} \right|,
    \\[0.2cm]
    \label{eq:MaxEntCond2}
    {\Delta x_j^\omega}^\alpha = \left| \bar{x}_j^\alpha - x_j'^\alpha \right| &=& \left| \tr\{X_j^\alpha(\bar\rho - \omega)\} \right|,
\end{IEEEeqnarray}
$j\in\{1,...,d_T\}$, between the true dynamical equilibrium values $\{\bar E_\alpha\,\{\bar x_j^\alpha\}\}$ and those associated with the globally constrained maximum entropy ensemble $\omega$: $\{E_\alpha^\omega,\{{x_j^\omega}^\alpha\}\}$, $\alpha=A$ or $B$.
The smaller these differences are, the closer $\bar S$ is to $S_\omega$ and, hence, to $S_\mathrm{max}$. 

If we demand that, for $N\to\infty$,
\begin{equation}\label{eq:SmallDeltaPrimes}
  \frac{\Delta E_A^\omega}{N} \to 0,\quad \frac{\Delta {x_j^\omega}^A}{N} \to 0,  
\end{equation}
that is, that these differences per particle become sufficiently small as $N$ increases, then the difference $|S_\omega-\bar S|/N$ vanishes asymptotically:
\begin{equation}\label{eq:maxentatequilib}
    \frac{|S_\omega - \bar S|}{N} \to 0.
\end{equation}
As a consequence, we also have $(S_\mathrm{max} - \bar S)/N \rightarrow 0$ when $N\to\infty$. 

In this case, since $\bar{S}$ becomes effectively indistinguishable from the upper bound $S_\mathrm{max}$, and since $S(t)$ is indistinguishable from $\bar S$ at almost all times, we can say $S(t)$ is effectively at maximum at almost all times.
The situation is illustrated in Fig.~\ref{fig:figIV}.
As a consequence of the effective maximization of $S(t)$, we have that~\eqref{eq:equiconds} holds at almost all times.
Notably, this implies that the inverse temperatures $\beta_A$ and $\beta_B$ of subsystems $A$ and $B$, as well as the other conjugate variables $\lambda_A$ and $\lambda_B$, equilibrate to effectively the same values.

In fact, the differences $\Delta E_A^\omega$ and ${\Delta x_j^\omega}^A$ may be seen as a metric of the ability of an observer measuring $\{E_\alpha,\{x_j^\alpha\}\}$ to distinguish between the states $\bar\rho$ and $\omega$.
When this distinguishability between the dynamical equilibrium state $\bar\rho$ and $\omega$ becomes sufficiently small, the system is said to thermalize from the perspective of this observer~\cite{Gogolin2016}.

A common justification for the thermalization of macroscopic systems is the Eigenstate Thermalization Hypothesis (ETH)~\cite{Jensen1985,Srednicki1994,Deutsch1991,Tasaki1998,Rigol2008,Gogolin2016}.
Let us consider the case where $E$ is the only thermodynamically relevant conserved quantity and let $|E_k\rangle$ denote an eigenstate of $H$ with eigenenergy $E_k$: $H|E_k\rangle=E_k|E_k\rangle$.
In one of its many forms, a system is said to satisfy the ETH if for any operator $O_A$, acting on subsystem $A$, $\langle E_k|O_A|E_k\rangle \approx \tr\{O_Ae^{-\beta(E_k) H_A}\}/\tr\{e^{-\beta(E_k) H_A}\}$~\cite{Tasaki1998}.
When this is the case, the condition $\Delta E_A^\omega \approx 0$ is necessarily satisfied.

From a different perspective, since Sec.~\ref{sec:EquiOfThermalVariables} we have been relying on the assumption that $E$ and $\{x_j\}$ constitute the only thermodynamically relevant quantities of the system.
Again, such assumption is based on the empirical fact that the equilibrium and thermodynamics of isolated macroscopic systems can be completely characterized by a small number of conserved quantities.
We can now see that the smallness of $\Delta E_A^\omega$ and ${\Delta x_j^\omega}^A$, as in~\eqref{eq:SmallDeltaPrimes}, is a necessary condition for the validity of this assumption.

Indeed, the dynamical equilibrium state $\bar\rho$ is nothing but the maximum von Neumann entropy ensemble constrained not only by $\{E,\,\{x_j\}\}$, but by the set of $d$ independent conserved quantities of the system, namely the probabilities of each energy eigenstate ~\cite{Gogolin2016,Sirker2014}. Hence, for a generic observable $O$, the difference in expected values $\tr\{O\bar\rho\}-\tr\{O\omega\}$ conveys information about conserved quantities other than $E$ and $\{x_j\}$~\cite{Sirker2014} --- see also Appendix~\ref{AppIII}.
As a consequence, if the differences $\Delta E_\alpha^\omega/N = |\tr\{H_\alpha (\bar\rho - \omega)\} |/N$ and $\Delta {x_j^\omega}^\alpha = | \tr\{ X_j^\alpha (\bar\rho - \omega) \} |/N$ are not sufficiently small, the ensemble $\omega$ --- constructed from $E$ and $\{x_j\}$ --- fails to reproduce the assumed measurable properties of $\bar\rho$ --- i.e., the equilibrium values $\{E_\alpha,\{x_j^\alpha\}\}$.
That is, if $\Delta E_\alpha^\omega/N$ and ${\Delta x_j^\omega}^\alpha/N$ are not sufficiently small, we can conclude that the equilibrium values $\{\bar{E}_\alpha,\{\bar{x}_j^\alpha\}\}$ of the subsystems depend significantly on quantities other than the energy $E$ and $\{x_j\}$.
Hence, in this case, $E$ and $\{x_j\}$ cannot be considered thermodynamically sufficient to completely characterize the system.

In~\cite{Sirker2014}, what highlights $E$ and $\{x_j\}$ in comparison with other conserved quantities is the local character of the associated observables $H$ and $\{X_j\}$; namely that in the thermodynamic limit they can be expressed as sums (or integrals) of operators with bounded spatial support.
Put differently, then, what the argument leading to~\eqref{eq:maxentatequilib} shows is: when the subsystems' equilibrium values $\{\bar{E}_\alpha,\{\bar x_j^\alpha\}\}$ depend solely on the thermodynamic-limit local conserved quantities $\{E,\{x_j\}\}$ and not on other nonlocal conserved quantities, then the entropy $S(t)$, defined in~\eqref{eq:bipentropy}, dynamically equilibrates to the maximum entropy $S_\omega$ constrained by $\{E,\{x_j\}\}$.

Our results are closely related to the recent work~\cite{Schindler2025}, which discusses the emergence of second laws as a consequence of the equilibration of quantum systems.
There, Jaynes’ maximum entropy principle is used to define a generalized observational entropy \(S_M^\omega(\rho) = -\sum_x p_x \ln p_x/V_x\), with $M=\{M_x\}$ a POVM, $p_x=\tr\{\rho M_x\}$, $V_x=e^{S(\omega)}\tr\{M_x\omega\}$ and $\omega$ the maximum entropy ensemble compatible with a given set of constraints.
It is then showed that if, to the observer measuring $M$, the system state $\rho(t)$ is close, on average, to $\omega$, the entropy $S_M^\omega(\rho(t))$ equilibrates on average to its maximal possible value, $S(\omega)$.
Furthermore, for two weakly coupled systems, when $M$ is a coarse measurement of the local energy in each subsystem, $S_M^\omega$ acquires a thermodynamic character and its maximization leads to equality of the subsystems’ Boltzmann temperatures.

While our approach shares conceptual similarities with Ref.~\cite{Schindler2025}, the two frameworks differ both in formulation -- involving different entropy functions -- and in the precise assumptions employed.
In both cases, entropy maximization emerges from an effective indistinguishability between the dynamical equilibrium state $\bar\rho$ and a maximum entropy ensemble $\omega$ associated with a restricted set of constraints.
However, the notion of distinguishability adopted is different.
In Ref.~\cite{Schindler2025}, the relevant condition is that the classical relative entropy \(D_M(\bar\rho||\omega)=\sum_x \bar p_x \ln \bar p_x/w_x\), with $\bar p_x=\tr\{M_x \bar\rho\}$ and $w_x=\tr\{M_x \omega\}$, associated with the measurement $M$, is small.
By contrast, in our framework, the assumption is that the expectation value differences density \(|\tr\{O_\alpha(\bar\rho - \omega)\}|/N\) are small, for a set of subsystem's experimentally realistic observables $O_\alpha$ related to the global constraints/conserved quantities.
We further argued that this effective indistinguishability is a necessary condition for the assumption that the equilibrium properties of the system depend only on the restricted set of conserved quantities used in the construction of $\omega$.

The role of coarse-graining also appears differently in the two approaches.
In Ref.~\cite{Schindler2025}, the coarse-graining must be encoded in the POVM $M$ defining the generalized observational entropy.
Here, this role is played by the finite resolution of experimentally realistic observables.


\section{\label{sec:ToyModel}Example}

To illustrate the equilibration of the entropies~\eqref{eq:entA},~\eqref{eq:entB} and~\eqref{eq:bipentropy}, and of the entropy conjugate variables~\eqref{eq:betadef} and~\eqref{eq:lambdef} as introduced above, let us consider the one-dimensional anisotropic Heisenberg model
\begin{equation}\label{eq:toymodel}
 \begin{aligned}
    H(N,\Delta,J_2) = &- J\sum_{j=1}^{N-1} (S_j^x S_{j+1}^x + S_j^y S_{j+1}^y + \Delta S_j^z S_{j+1}^z) 
    \\[0.2cm]
    &- J_2\sum_{j=1}^{N-2} (S_j^x S_{j+2}^x + S_j^y S_{j+2}^y + \Delta S_j^z S_{j+2}^z),
 \end{aligned}
\end{equation}
where $S_j^\alpha$ are spin-1/2 operators at site $j$, $J\equiv1$ and $J_2$ are the couplings between nearest and next-nearest neighbors, respectively, and $\Delta$ gives the anisotropy of the interaction.
For $J_2\neq0$, this model is nonintegrable, with $H$ itself and the $z$-component of total spin, $S^z=\sum_j S_j^z$, being the only conserved local operators, whose expected values we assume to be the system's thermodynamically relevant conserved quantities.
\begin{figure}
    \centering
    \includegraphics[width=1.0\linewidth]{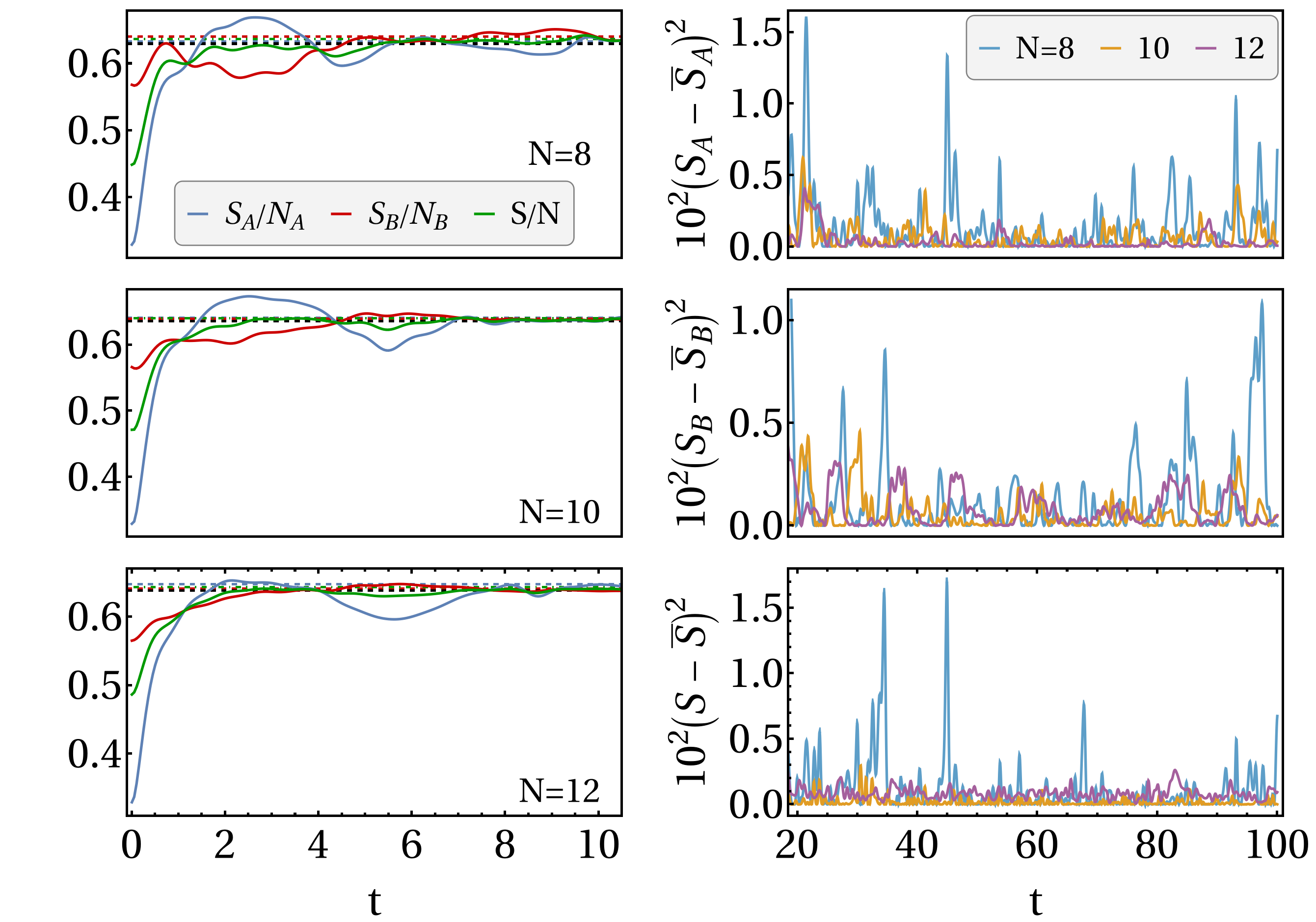}
    \caption{\label{fig:EntrGrid} \emph{Left:} The behaviors of $S_A/N_A$, $S_B/N_B$ and $S/N$ as functions of time for different system sizes $N$.
    Their respective equilibrium values are shown by the colored dashed lines.
    The dashed black lines also show the values of the entropy $S_\omega/N$ in~\eqref{eq:maxentprime} associated with the global constraints $E = \langle \psi|H|\psi\rangle$ and $m = \langle\psi|S^z|\psi\rangle$.
    \emph{Right:} Squared differences between the entropies $S_A$ (top), $S_B$ (middle), and $S$ (bottom) and their respective equilibrium values for different system sizes.}
\end{figure}

In what follows we divide the chain of $N=N_A+N_B$ spins into two parts $A$ and $B$ with $N_A$ and $N_B$ spins, respectively.
Thus, let $H_\alpha = H(N_\alpha,\Delta,J_2)$, $\alpha=A, B$, denote the Hamiltonian of subsystem $\alpha$, and $S_A^z=\sum_{j=1}^{N_A} S_j^z$ and $S_B^z=S^z-S_A^z$ denote the $z$-components of total spin in subsystems $A$ and $B$, respectively.
We fix $N_A=4$, $J_2=0.2$ and $\Delta=0.8$ and consider initial states of the form
\begin{equation}
    \begin{aligned}
        |\psi\rangle &= |\psi_A\rangle\otimes|\psi_B\rangle,
        \\[0.2cm]
        |\psi_\alpha\rangle &= \sum_k \frac{ \exp[-\frac{1}{2}(\beta_\alpha^0 e_k^\alpha + \lambda_\alpha^0 m_k^\alpha)]}{\sqrt{Z_\alpha^0}}e^{i\phi_k^\alpha}|k_\alpha\rangle,
    \end{aligned}
\end{equation}
where $\phi_k^\alpha$ are random phases, $\{|k_\alpha\rangle\}$ are eigenvectors of $H$ and $S_\alpha^z$ with eigenvalues $\{e_k^\alpha\}$ and $\{m_k^\alpha\}$, respectively, and $Z_\alpha^0=\sum_k\{\exp(-\beta_\alpha^0 e_k^\alpha - \lambda_\alpha^0 m_k^\alpha)\}$.
In such states, the initial values of inverse temperature $\beta_\alpha^0$ and field $\lambda_\alpha^0$ of each subsystem is fixed.
Specifically, in all cases, we take $\beta_A^0=2$, $\beta_B^0=0.05$ and $\lambda_A^0=-\lambda_B^0=0.5$.

The left panels of Figure~\ref{fig:EntrGrid} show the behaviors of the entropies per particle $S_A/N_A$ and $S_B/N_B$, of subsystems $A$ and $B$, as well as the total entropy $S/N=(S_A + S_B)/N$, as functions of time for different total number os spins $N$.
In all cases we see a swift convergence of these quantities to their respective equilibrium values, $\bar{S}_A/N_A$, $\bar{S}_B/N_B$ and $\bar{S}/N$, indicated by the colored dashed lines.
Notably, in all cases, the equilibrium value of $S$ is also close to $S_\omega/N$ -- represented by the black dashed lines -- which is the entropy in~\eqref{eq:maxentprime} associated with the global fixed constraints on the energy $E=\langle\psi|H|\psi\rangle$ and $z$-component of total magnetization $m =\langle \psi|S^z|\psi\rangle$.
The right side of Fig.~\ref{fig:EntrGrid} also shows the time-behavior of the square of the differences between these entropies and their respective equilibrium values --- namely, $(S_A - \bar{S}_A)^2$, $(S_B - \bar{S}_B)^2$ and $(S-\bar{S})^2$ --- for different system sizes, after some transient.
These confirm a decrease in the fluctuations of $S_A$, $S_B$ and $S$ around their equilibrium values as the total size of the system increases.
\begin{figure}
    \centering
    \includegraphics[width=1.0\linewidth]{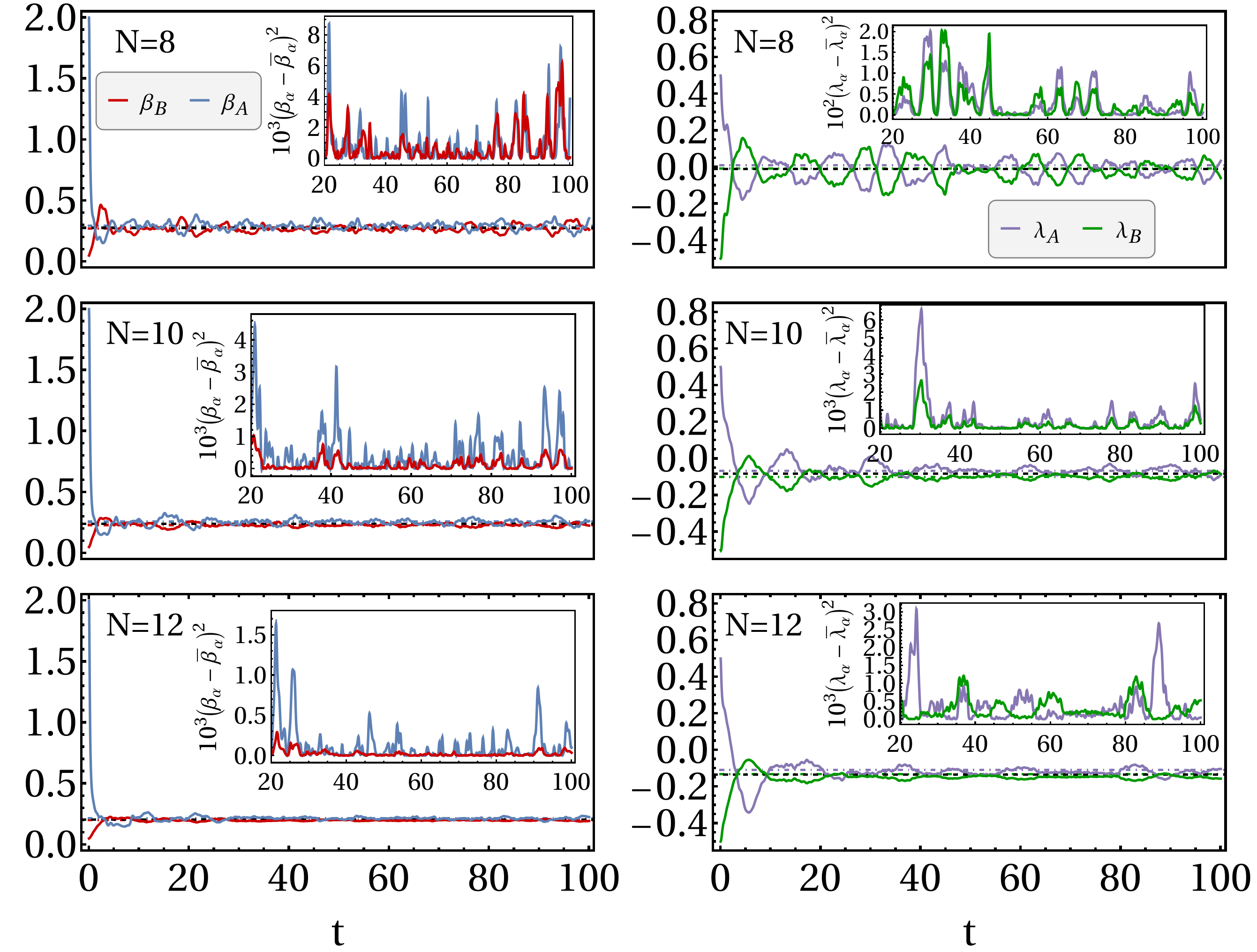}
    \caption{\label{fig:LagrGrid} \emph{Left:} The behaviors of the inverse temperatures $\beta_A$ and $\beta_B$ of subsystems $A$ and $B$ as functions of time for different system sizes $N$.
    The colored dot-dashed lines show their respective equilibrium values while the black dashed lines show the values of $\beta$ in~\eqref{eq:maxentprime} associated with the global constraints $E = \langle \psi|H|\psi\rangle$ and $m = \langle\psi|S^z|\psi\rangle$.
    The insets also show the the squared differences between $\beta_A$, $\beta_B$ and their respective equilibrium values as functions of time after an initial transient time.
    \emph{Right:} Equivalent plots for the entropy conjugate variables of the $z$-component of the total spin in each subsystem.}
\end{figure}

Figure~\ref{fig:LagrGrid} similarly shows the behaviors of the inverse temperatures $\beta_A$ and $\beta_B$, of subsystems $A$ and $B$, as well as of $\lambda_A$ and $\lambda_B$, as functions of time for different system sizes.
Specifically, the left panels show that $\beta_\alpha$, $\alpha=A,\,B$, quickly converges to points near its respective equilibrium value $\bar{\beta}_\alpha$, indicated by a colored dot-dashed line.
Notice that these are also close to the value of the inverse temperature $\beta$ associated with the global constraints $E=\langle\psi|H|\psi\rangle$ and $m=\langle\psi|S^z|\psi\rangle$ in~\eqref{eq:maxentprime}.
The insets in these panels further show the behavior of the fluctuations of $\beta_\alpha$, $\alpha=A,\,B$, near their equilibrium values as captured by the squared differences $(\beta_\alpha - \bar{\beta}_\alpha)^2$, after some transient time.
In particular, they show a decrease in these fluctuations as the total system size increases.
The right panels show analogous results for $\lambda_A$ and $\lambda_B$.
Namely, that $\lambda_\alpha$, $\alpha=A,B$, converge to its equilibrium value $\bar{\lambda}_\alpha$, represented by colored dot-dashed lines in the figure, which is also close to $\lambda$ -- black dashed lines -- obtained from the global constraints $E$ and $m$.
Moreover, the insets also show the decrease in the fluctuations $(\lambda_\alpha - \bar{\lambda}_\alpha)^2$ as the size of the system increases.

It is important to notice that for such small systems, all bounds derived in Sec.~\ref{sec:EquiOfThermalVariables}, namely Eqs.~\eqref{eq:betaBDeq},~\eqref{eq:lambBDeq} and~\eqref{eq:entABBD} and~\eqref{eq:entBD}, are loose.
Yet, we can still see the same qualitative behaviors expected to hold for truly macroscopic systems of the entropies and entropy conjugated variables defined here.
Most significantly on Figures~\ref{fig:EntrGrid} and~\ref{fig:LagrGrid} is the stronger convergence to the expected behaviors of these quantities as the total size of the system is increased.


\section{\label{sec:conc}Conclusion}

Determining how thermodynamic equilibrium conditions emerge from the microscopic quantum dynamics of isolated systems is a central problem in the foundations of statistical mechanics.
While important progress has been achieved regarding the equilibration of expectation values of observables, thermodynamics also involves quantities that are not themselves observables in the quantum mechanical sense.

In this work we showed that any continuously differentiable function of equilibrating expectation values also equilibrates on average.
More precisely, we derived a bound on the time-averaged squared fluctuations of such functions, which becomes exponentially small in systems with exponentially dense spectra.
This allowed us to extend equilibration-on-average arguments from quantum observables to thermodynamic quantities defined as functions of equilibrating expectation values.

Considering two weakly coupled subsystems exchanging energy and other globally conserved extensive quantities, we studied thermodynamic variables defined as functions of the local values of the exchanged quantities through Jaynes’ maximum entropy principle.
In particular, we showed the equilibration of entropy, temperature and other entropy-conjugate variables that together form a thermodynamic fundamental relation for the subsystems.
We further showed that, when the local equilibrium values of the exchanged quantities depend solely on the global conserved quantities, the equilibration of the total entropy occurs effectively at maximum and leads to the usual thermodynamic equilibrium condition of equality of entropy-conjugate variables across the subsystems.

Several extensions to the present work deserve further investigation.
An important example concerns thermodynamic variables such as volume, which play a central role in the description of gases, but which are not naturally represented as expectation values of quantum observables.
It would therefore be interesting to understand how the present approach could be generalized to incorporate variables associated with boundary or countor conditions.

Another key direction concerns the investigation of the time scales involved in the emergence of thermodynamic behavior~\cite{Wilming2018,de_Oliveira_2018}.
In particular, the perspective of dynamical equilibration as the root for an effective second law of thermodynamics is only meaningful when equilibration occurs on time scales much shorter than those associated with recurrences on the system state.
Obtaining more explicit relations between equilibration and recurrence times would therefore help clarify the regime in which the present thermodynamic picture applies.

Finally, both the present work and related approaches~\cite{Schindler2025} suggest that entropy maximization relies on the dominant role played, in the average dynamics of large systems, by local conserved quantities over nonlocal ones.
Understanding the mechanisms responsible for this distinction remains an interesting direction for future research.

\section*{Acknowledgments}
This work is supported by the National Council for Scientific and Technological Development, CNPq Brazil (projects: Universal Grant No. 408990/2025-2, and 409611/2022-0). TRO acknowledges funding from the Air Force Office of Scientific Research under Grant No. FA9550-23-1-0092.


\appendix

\section{\label{AppI} Maximum Entropy Principle}

Let $\{O_j\}_{j=1}^n$ be a set of bounded and linearly independent observables used to probe a quantum system.
We denote by 
\[
    \mathcal{R} = \{(o_1,\dots,o_n)\in\mathbb{R}^n\,|\,\exists\,\rho\ge0,\, \tr\{\rho\}=1,\, o_j=\tr\{O_j\rho\} \}
\]
the convex set of physically acceptable expectation-value vectors $\mathbf{o}=(o_1,\dots,o_n)$ for this set of observables~\cite{Wichmann1963}.

Let us assume the only available information about this system is encoded in a vector $\mathbf{o}=(o_1,\cdots,o_n)$ of expectation values.
The maximum entropy principle states that the least biased description of the system consistent with this information is provided by the ensemble $\varsigma$ maximizing the von Neumann entropy~\cite{Jaynes1957}
\[
 S(\rho) = -\tr\{\rho\ln\rho\},
\]
subject to the constraints
\[
 \tr\{O_j\varsigma\} = o_j,
\]
for $j\in\{1,\dots,n\}$.

In particular, if $\mathbf{o}$ lies in the interior of $\mathcal{R}$, the \emph{unique} solution to this optimization problem is given by~\cite{Wichmann1963,Ingarden1997}
\[
 \begin{aligned}
 \varsigma &= \frac{1}{\mathcal{Z}}\exp(-\sum_{j=1}^n \lambda_j O_j),
 \\[0.2cm]
 \mathcal{Z} &= \tr\left\{\exp(-\sum_{j=1}^n \lambda_j O_j)\right\},
 \end{aligned}
\]
and the associated (maximum) entropy reads
\begin{equation}\label{eq:entJaynes}
    S(\varsigma) = \ln \mathcal{Z} + \sum_{j=1}^n \lambda_j o_j.
\end{equation}
The finite Lagrange multipliers $\lambda_j$ are determined by the one-to-one map of constraints~\cite{Wichmann1963}
\[
 o_j = \tr\{O_j\varsigma\} = - \frac{\partial \ln\mathcal{Z} }{\partial \lambda_j}.
\]
This means we can invert this set of constraint equations to obtain $\lambda_j(\mathbf{o})$.

In particular, if all $O_j$ are experimentally realistic and have exponentially dense spectra --- that is, exponentially close eigenspaces --- then all experimentally accessible expected-value vectors $\mathbf{o}$ must lie in the interior of $\mathcal{R}$. 
To illustrate this point, let us consider first the case of a single observable $O$.
Then, $\mathcal{R}=[\lambda_\mathrm{min}(O),\,\lambda_\mathrm{max}(O)]$ is the closed interval delimited by the smallest and largest eigenvalues $\lambda(O)$ of $O$.
A system state realizing one of these boundary values corresponds to a state supported entirely on the eigenspace associated with a single eigenvalue of $O$.
But, as discussed in Section~\ref{sec:EquiOfRealObs}, the preparation of a state with any realistic uncertainty will surely lead to the occupation of a gigantic number of the exponentially close eigenspaces of $O$.

Similarly, for the set of operators $\{O_j\}_{j=1}^n$, the boundary points of $\mathcal{R}$ are given by extremal eigenspaces of $O_u=\sum_j u_jY_j$, where $\mathbf{u}=(u_1,...,u_n)$ is a unit vector in $\mathbb{R}^n$~\cite{Wichmann1963}.
If all $O_j$ have exponentially dense spectra, so does $O_\mathbf{u}$ and the previous argument follows straightforwardly.
As a consequence, states associated with exact boundary points in $\mathcal{R}$ are operationally inaccessible.

The effective exclusion of this boundary is further reinforced by the finite resolution of experimentally realistic observables.
With finite resolution $\delta O$, points in $\mathcal{R}$ differing by less than $\delta O$ are in essence indistinguishable; the experimental information, therefore, lacks the necessary precision to justify assigning to $o=\tr\{O\rho\}$ a value identical to a boundary point.
Even if close to $\lambda_\mathrm{min}(O)$ or $\lambda_\mathrm{max}(O)$, the experimentally determined expected-value $o$ must necessarily lie inside their delimited interval.

In Sec.~\ref{sec:EquiOfThermalVariables} we consider a set $\{H_\alpha, \{X_j^\alpha\}\}$ of extensive -- implying exponentially dense spectra -- and experimentally realistic observables.
Together with a realistic state preparation $\rho_0$, this ensures the expected values $\{E_\alpha(t),\{x_j^\alpha(t)\}\}$ lie in the interior of the set of constraints $\mathcal{R}(\{H_\alpha,\{x_j^\alpha\}\})$.
At each time $t$, then, we may construct the maximum-entropy ensemble
\begin{equation}\label{eq:maxensalpha}
  \sigma_\alpha^t = \frac{1}{Z_\alpha(t)}\exp(-\beta_\alpha(t) H_\alpha - \sum_{j=1}^{d_T}\lambda_j^\alpha(t) X_j^\alpha),   
\end{equation}
\[
 \begin{aligned}
 Z_\alpha(t) &= \tr\left\{\exp(-\beta_\alpha(t) H_\alpha - \sum_{j=1}^{d_T}\lambda_j^\alpha(t) X_j^\alpha)\right\},
 \\[0.2cm]
 E_\alpha(t) &= -\frac{\partial \ln Z_\alpha(t)}{\partial \beta_\alpha(t)}, \quad x_j^\alpha(t) = -\frac{\partial \ln Z_\alpha(t)}{\partial \lambda_j^\alpha(t)},
 \end{aligned}
\]
whose entropy
\[
 S(\sigma_\alpha^t) = \ln Z_\alpha(t) + \beta_\alpha(t)E_\alpha(t) + \sum_{j=1}^{d_T}\lambda_j^\alpha(t)x_j^\alpha(t)
\]
we consider as the thermodynamic entropy of subsystem $\alpha$.


\section{\label{AppCo} Covariance Matrix}

Let $\lambda_{0}^\alpha\equiv \beta_\alpha$, $x_{0}^\alpha\equiv E_\alpha$ and $X_{0}^\alpha \equiv H_\alpha$.
The invertibility of equations~\eqref{eq:noneqTemp} and~\eqref{eq:noneqXjs} means there exist differentiable functions $f_j^\alpha$ such that
\[
 \lambda_j^\alpha = f_j^\alpha \left(\{x_k^\alpha\}_{k=0}^{d_T} \right).
\]
To establish the bounds~\eqref{eq:betaBDeq} and~\eqref{eq:lambBDeq}, based on the mean value theorem, we must compute the derivatives $\partial f_j^\alpha/\partial x_k^\alpha$.
But since we do not have an explicit form for $f_j^\alpha$, we want to express these derivatives in terms of $Z_\alpha$.

Therefore, let $J_\alpha$ denote the matrix whose elements are precisely $\partial f_j^\alpha/\partial x_k^\alpha$,
\[
    [J_\alpha]_{jk} = \frac{\partial \lambda_j^\alpha}{\partial x_k^\alpha} \equiv \frac{\partial f_j^\alpha}{\partial x_k^\alpha}.
\]

Let also $C_\alpha$ denote the matrix with elements
\[
    [C_\alpha]_{jk} = \frac{\partial x_j^\alpha}{\partial \lambda_k^\alpha}.
\]
It is straightforward to check that 
\[
 J_\alpha C_\alpha = \mathds{1} \Rightarrow J_\alpha = C_\alpha^{-1},
\]
where $\mathds{1}$ is the identity matrix and $C_\alpha^{-1}$ the inverse of $C_\alpha$.
 Thus, 
\begin{equation}\label{eq:invcov}
 \frac{\partial f_j^\alpha}{\partial x_k^\alpha} = \left[ C_\alpha^{-1} \right]_{jk}.    
\end{equation}
These derivatives can also be written as
\[
 \frac{\partial f_j^\alpha}{\partial x_k^\alpha} =  \frac{(-1)^{j+k} \det C_\alpha^{(k,\,j)}}{\det C_\alpha},
\]
with $C_\alpha^{(k,\,j)}$ the submatrix obtained from $C_\alpha$ by removing row $k$ and column $j$.

As derived in Eq.~\eqref{eq:maxensalpha}, for the maximum entropy ensemble $\sigma_\alpha^t$ constrained by the expected values $x^\alpha_i$, the latter satisfy  $x_i^\alpha = -\partial \ln Z_\alpha/\partial \lambda_i^\alpha$. Thus, $C_\alpha$ is a positive-definite matrix~\cite{Wichmann1963} with elements
\begin{align*}
 \left[C_\alpha(t) \right]_{jk} & = -\frac{\partial^2 \ln Z_\alpha}{\partial \lambda_k^\alpha \partial \lambda_j^\alpha} \\
    &= \tr\{X_j^\alpha X_k^\alpha\sigma_\alpha^t\} - \tr\{X_j^\alpha\sigma_\alpha^t\}\tr\{X_k^\alpha\sigma_\alpha^t\}, 
\end{align*}
which are the covariances of the observables $X_k^\alpha$ and $X_j^\alpha$ in the state $\sigma_\alpha^t$.
That is, $[C_\alpha]_{jk}$ quantifies the correlations between fluctuations of $X_j^\alpha$ and $X_k^\alpha$ in the thermal state $\sigma_\alpha$.

Away from phase-transition points, thermal correlations are short-ranged and the covariances of extensive thermal quantities are also extensive, $[C_\alpha]_{ij}\sim N$.
Given that $\{E_\alpha,\{x_j^\alpha\}\}$ constitute independent thermodynamic coordinates, the covariance matrix $C_\alpha$ remains nonsingular in the thermodynamic limit.
Since its elements scale linearly with system size, the elements of its inverse scale as $N^{-1}$.
Therefore,
\[
 \frac{\partial f_j^\alpha}{\partial x_k^\alpha} \sim \frac{1}{N}.
\]

In eqs.~\eqref{eq:betaBDeq} and~\eqref{eq:lambBDeq}, $\mu_\alpha \equiv \mu_{\alpha\, j=0}$ and $\mu_{\alpha\, j}$ are the maximum absolute values of $\partial f_0^\alpha/\partial x_k^\alpha$ and $\partial f_j^\alpha/\partial x_k^\alpha$, respectively, computed at the mean-value points connecting $\lambda_j^\alpha(t)$ to its equilibrium value $\bar \lambda_j^\alpha$.
Hence, whenever $[C_\alpha]_{jk}\sim N$,
\[
 \mu_{\alpha j}\sim 1/N.
\]



\section{\label{AppAltBounds}Alternative Bounds on $S_\alpha$}

From the concavity of $S_\alpha$ it follows that
\[
 L \le S_\alpha(t) - \bar S_\alpha \le R,
\]
where
\[
 \begin{aligned}
     L & = \beta_\alpha(t) \left( E_\alpha(t) - \bar E_\alpha \right) + \sum_{j=1}^{d_T} \lambda_j^\alpha(t) \left( x_j^\alpha(t) - \bar x_j^\alpha \right),
     \\[0.2cm]
     R & = \bar{\beta}_\alpha \left( E_\alpha(t) - \bar E_\alpha \right) + \sum_{j=1}^{d_T} \bar\lambda_j^\alpha \left( x_j^\alpha(t) - \bar x_j^\alpha \right).
 \end{aligned}
\]

Hence,
\[
\begin{aligned}
    \left( S_\alpha(t) - \bar{S}_\alpha \right)^2 &\le \max\{ L^2,\, R^2\}
    \\[0.2cm]
    & \le \max\{\bar \lambda_\alpha^2,\, {\lambda_\alpha'}^2\} \left(\sum_{j=0}^{d_T} \left| x_j^\alpha(t) - \bar x_j^\alpha \right| \right)^2,
\end{aligned}
\]
where $|x_0^\alpha(t) - \bar x_0^\alpha| = |E_\alpha(t) - \bar E_\alpha|$ and
\[
 \begin{aligned}
 \bar \lambda_\alpha &= \max \left\{ \left|\bar \lambda_j^\alpha \right| \right\}_{j=0}^{d_T},
 \\[0.2cm]
 \lambda_\alpha' &= \max_{\displaystyle t\in [0,\tau]} \max \left\{ \left|\lambda_j^\alpha(t) \right| \right\}_{j=0}^{d_T},
 \end{aligned}
\]
where $\bar\lambda_{j=0}^\alpha=\bar \beta_\alpha$ and $\lambda_{j=0}^\alpha(t) = \beta_\alpha(t)$.
Notice that $\lambda_\alpha'$ differs from $\lambda_\alpha$ in~\eqref{eq:betaBDeq} by the fact that the maximization over time in the former takes into account the actual evolution of the multipliers $\{\lambda_j^\alpha(t)\}_{j=0}^{d_T}$ while the latter takes into account their values at the mean-value points $\left\{\boldsymbol{\xi}_t^{\, S_\alpha}\right\}$.

Averaging over time we obtain
\[
 \overline{\left( S_\alpha(t) - \bar S_\alpha \right)^2}^{\,\tau} \le \max\{ \bar \lambda_\alpha^2,\, {\lambda_\alpha'}^2\} \left(||H_\alpha|| + \sum_{j=1}^{d_T}||X_j^\alpha||\right)^2\varepsilon.
\]

Similarly,
\[
    \overline{(S(t)-\bar{S})^2}^{\,\tau} \le \sum_{\alpha=A,B} \max\{\bar\lambda_\alpha^2,\,{\lambda'}_\alpha^2\} \left(||H_\alpha|| + \sum_{j=1}^{d_T}||X_j^\alpha||\right)^2\varepsilon.
\]


\section{\label{AppIII}Total Entropy Asymptotic Bounds}

As in Sec.~\ref{sec:EquiToMaxEnt} we denote by
\[
 \omega = \exp(-\beta H - \sum_{j=1}^{d_T}\lambda_j X_j)/Z,
\]
the entropy maximizing ensemble constrained by the conserved quantities $\{E,\{x_j\}\}$ and by
\[
 \sigma_\mathrm{max} = \frac{e^{-\beta_\mathrm{max} (H_A + H_B) + \sum_{j=1}^{d_T}\lambda_j^\mathrm{max} (X_j^A+X_j^B)}}{Z_A^\mathrm{max}Z_B^\mathrm{max}},
\]
the ensemble maximizing the thermodynamic entropy $S$~\eqref{eq:bipentropy}.
Specifically, $\sigma_\mathrm{max}$ is the entropy maximizing ensemble constrained by the subsystems' values $\{E_A, \{x_j^A\}\}$ and $\{E_B \approx E-E_A,\,\{x_j^B = x_j - x_j^A\}\}$ that gives maximum $S$: $S_\mathrm{max}=S(\sigma_\mathrm{max})$.
From~\eqref{eq:textbookEquilibEq} we know that at this maximum $\beta_A = \beta_B \equiv\beta_\mathrm{max}$ and $\lambda_A = \lambda_B \equiv \lambda_\mathrm{max}$.

Let $S(\rho||\sigma)=\tr\{\rho(\ln \rho - \ln \sigma)\}\ge0$ denote the relative entropy between states $\rho$ and $\sigma$.
From $S(\sigma_\mathrm{max}||\omega) \ge 0$ we get
\begin{equation}
    \beta (E - \tr\{H\sigma_\mathrm{max}\}) + \sum_{j=1}^{d_T}\lambda_j(x_j - \tr\{X_j\sigma_\mathrm{max}\}) \le S_\omega - S_\mathrm{max},
\end{equation}
where we used Eq.~\eqref{eq:maxentprime}.
Conversely, from $S(\omega||\sigma_\mathrm{max})\ge0$ we obtain
\begin{equation}
 \begin{aligned}
     S_\omega - S_\mathrm{max} \le \beta_\mathrm{max} &  \tr\{(H_A+H_B)(\omega - \sigma_\mathrm{max})\} 
    \\[0.2cm]
      + \sum_{j=1}^{d_T} & \lambda_j^\mathrm{max} \tr\{(X_j^A + X_j^B)(\omega - \sigma_\mathrm{max})\}.
 \end{aligned}
\end{equation}

Now, by construction $\sigma_\mathrm{max}$ is such that
\[
\tr\{X_j\sigma_\mathrm{max}\} = \tr\{(X_j^A + X_j^B)\sigma_\mathrm{max}\} = x_j = \tr\{X\omega\}.
\] 
As a consequence, the $x_j$-depend terms in the previous expressions vanish.
Moreover, $\sigma_\mathrm{max}$ must be such that $\tr\{(H_A+H_B)\sigma_\mathrm{max}\}$ differs from $E=\tr\{H\omega\}$ by at most something of the order of $||H_I||$.
Thus,
\[
 \begin{aligned}
     |E \! - \! \tr\{H\sigma_\mathrm{max}\}| &= |E \! - \! \tr\{(H_A+H_B)\sigma_\mathrm{max}\} \! - \!  \tr\{H_I\sigma_\mathrm{max}\}|
     \\[0.2cm]
     &\le |E \! - \! \tr\{(H_A+H_B)\sigma_\mathrm{max}\}|
     \!+ \! |\tr\{H_I\sigma_\mathrm{max}\}|
     \\[0.2cm]
     &\le 2||H_I||,
 \end{aligned} 
\]
and
\[
 \begin{aligned}
     |\tr\{(H_A\! + \! H_B)(\omega \! - \!\sigma_\mathrm{max})\}| &= |E \! - \!  \tr\{H_I\omega +(H_A \! + \!H_B)\sigma_\mathrm{max}\}|
     \\[0.2cm]
     &\le 2||H_I||,
 \end{aligned} 
\]
Hence, we have
\begin{equation}
     |S_\omega - S_\mathrm{max}| \le 2 ||H_I|| \max\{|\beta|, |\beta_\mathrm{max}|\}.
\end{equation}

Therefore, since $||H_I||/N\to0$, also
\begin{equation}
     \frac{1}{N}|S_\omega - S_\mathrm{max}|\to0.
\end{equation}
That is, $S_\omega$ and $S_\mathrm{max}$ become asymptotically equal as $N$ increases when the interaction $H_I$ is subextensive.

Meanwhile, let
\[
 \bar \sigma = \frac{e^{-\bar\beta_A H_A + \sum_{j=1}^{d_T}\bar\lambda_j^A X_j^A}}{\bar Z_A}\otimes \frac{e^{-\bar\beta_B H_B + \sum_{j=1}^{d_T}\bar\lambda_j^A X_j^B}}{\bar Z_B}
\]
denote the ensemble such that $\bar S = S(\bar \sigma)$, where $\bar S$ is the equilibrium value of $S$.
From $S(\bar\sigma||\omega)\ge0$, we get
\begin{equation}\label{eq:lowboundbarS}
    \beta(E - \tr\{H\bar\sigma\}) \le S_\omega - \bar S,
\end{equation}
where we used that $\bar x_j^A + \bar x_j^B = x_j$.
From $S(\omega||\bar\sigma)\ge0$, we get
\begin{equation}
\begin{aligned}
 S_\omega - \bar{S} &\le \bar\beta_A (E_A^\omega - \bar{E}_A) + \bar\beta_B (E_B^\omega - \bar{E}_B) 
 \\[0.2cm]
 &\quad+ \sum_{j=1}^{d_T} \bar\lambda_j^A({x_j^\omega}^A-\bar{x}_j^A) + \bar\lambda_j^B({x_j^\omega}^B - \bar{x}_j^B),
\end{aligned}
\end{equation}
where
\[
 \begin{aligned}
     \bar E_\alpha = \tr\{H_\alpha\bar\rho\} = \tr\{H_\alpha \bar\sigma\},
     \\[0.2cm]
     \bar x_j^\alpha  = \tr\{X_j^\alpha\bar\rho\} = \tr\{X_j^\alpha \bar\sigma\},
 \end{aligned}
\]
with the first equalities arising from the definition of equilibrium values and the second from the definition of $\bar\sigma$.
Moreover, $E_\alpha^\omega=\tr\{H_\alpha \omega\}$ and ${x_j^\omega}^\alpha = \tr\{X_j^\alpha \omega\}$.

Manifestly, ${x_j^\omega}^A + {x_j^\omega}^B = x_j = \bar{x}_j^A + \bar{x}_j^B$.
Furthermore,
\[
 \begin{aligned}
     E &= \tr\{H \bar\rho\} = \bar E_A + \bar E_B + \tr\{H_I \bar \rho\}
     \\[0.2cm]
     &= \tr\{H\omega\} = E_A^\omega + E_B^\omega + \tr\{H_I \omega\},
 \end{aligned}
\]
and thus $E_B^\omega - \bar E_B = -(E_A^\omega - \bar E_A) + \tr\{H_I(\bar\rho - \omega)\}$.
Therefore,
\begin{equation}
\begin{aligned}\label{eq:upboundbarS}
 S_\omega - \bar{S} &\le (\bar\beta_B - \bar\beta_A) (E_A^\omega - \bar{E}_A) + \bar\beta_B\tr\{H_I(\bar\rho - \omega)\} 
 \\[0.2cm]
 &\quad+ \sum_{j=1}^{d_T} (\bar \lambda_j^B - \bar\lambda_j^A) ({x_j^\omega}^A-\bar{x}_j^A).
\end{aligned}
\end{equation}

In~\eqref{eq:lowboundbarS},
\[
\begin{aligned}
 E - \tr\{H\bar\sigma\} = \tr\{H(\bar \rho - \bar\sigma)\} = \tr\{H_I(\bar \rho - \bar\sigma)\}.   
\end{aligned}
\]
Combining this with~\eqref{eq:upboundbarS},
\begin{equation}
 \begin{aligned}
    &|S_\omega - \bar S| \le \max \Bigg\{|\beta|\,|\tr\{H_I(\bar \rho - \bar\sigma)\}|,\,
    \\[0.6cm]
    &|\bar\beta_B|\,|\tr\{H_I(\bar\rho - \omega)\}| + |\bar \beta_B - \bar \beta_A||\tr\{H_A(\bar \rho - \omega)\}| 
    \\[0.2cm]
    &+ \sum_{j=1}^{d_T} |\bar \lambda_j^B - \bar \lambda_j^A||\tr\{X_j^A(\bar \rho - \omega)\}|
    \Bigg\}. 
 \end{aligned}
\end{equation}

Hence, using that $||H_I||/N\to0$, in the limit $N\to\infty$,
\begin{equation}
 \begin{aligned}
    \frac{|S_\omega - \bar S|}{N} &\le |\bar \beta_B - \bar \beta_A| \frac{|\tr\{H_A(\bar \rho - \omega)\}|}{N} 
    \\[0.2cm]
    &+ \sum_{j=1}^{d_T} |\bar \lambda_j^B - \bar \lambda_j^A| \frac{|\tr\{X_j^A(\bar \rho - \omega)\}|}{N}.
 \end{aligned}
\end{equation}
That is, in order to get
\[
 \frac{1}{N}|S_\omega - \bar S| \to0
\]
in the thermodynamic limit, we must have in this limit
\[
 \begin{aligned}
     \frac{1}{N} \left|\tr\{H_A(\bar\rho - \omega)\} \right| \to 0,
     \\[0.2cm]
     \frac{1}{N} \left|\tr\{X_j^A(\bar\rho - \omega)\} \right| \to 0.
 \end{aligned}
\]


\section{Thermodynamically Relevant and Irrelevant Conserved Quantities}

Since $E$ and $\{x_j\}$ are conserved quantities, their associated operators commute --- $[H,X_j]=[X_i,X_j]=0$, $i,j=1,...,d_T$ --- and, therefore, share a common eigenbasis.
Let $\{|k\rangle\}_{k=1}^d$ denote this eigenbasis and $P_k=|k\rangle\langle k|$.
The dynamical equilibrium state $\bar \rho$ consists of the maximum (von Neumann) entropy ensemble conforming to all constant occupations $p_k=\tr\{P_k \rho(t)\}=\tr\{P_k \rho(0)\}$~\cite{Gogolin2016}:
\begin{equation}\label{eq:maxentrhobar1}
 \bar \rho = \exp(-\sum_{k=1}^{d} \zeta_k P_k)/Z_\zeta,
\end{equation}
where $Z_\zeta=\tr\{\exp(-\sum_k\zeta_kP_k)\}=\sum_ke^{-\zeta_k}$, with $\zeta_k$ implicitly given by $p_k = -\partial_{\zeta_k}\ln Z_\zeta=e^{-\zeta_k}/Z_\zeta$.

By an orthogonal transformation, however, we can write $\bar\rho$ instead as~\cite{Sirker2014}
\begin{equation}\label{eq:maxentrhobar2}
    \bar \rho = \exp(- \bar\beta H - \sum_{j=1}^{d_T} \bar \lambda_j X_j - \sum_{k=d_T+1}^{d} \bar\zeta_k \bar P_k)/Z_\zeta,
\end{equation}
where $\bar\beta$ and $\bar \lambda_j$ are determined by $E = \tr\{H\bar\rho\} = - \partial_{\bar\beta} \ln Z_\zeta$ and $x_j = \tr\{X_j\bar\rho\} = -\partial_{\bar\lambda_j} \ln Z_\zeta$, and where $\{\bar{P}_k\}$ are orthogonal to $H$ and $\{X_j\}$ with respect to the Hilbert-Schmidt inner product weighted by $\omega$ (the maximum entropy ensemble constrained solely by $\{E,\{x_j\}\}$):
\begin{equation}\label{eq:thinnerprod}
 \tr\{\bar{P}_kH\omega\} = \tr\{\bar P_k X_j\omega\} = 0,
\end{equation}
for $k=d_T+1,\dots,d$ and $j=1,\dots, d_T$.

Specifically, $\{P_k\}$ provide a basis for all operators of the kind $K_\mathrm{diag} = \sum_k \kappa_k P_k$.
What the above transformation does is to define a new operator basis which allows the decomposition 
\[
K_\mathrm{diag} = K_{\{H,\{X_j\}\}} + K_{\{\bar{P}_k\}}
\]
where, in terms of the weighted inner product~\eqref{eq:thinnerprod}, $K_{\{H,\{X_j\}\}}$ belongs to the subspace spanned by $\{H,\{X_j\}\}_{\omega}$ and $K_{\{\bar P_k\}}$ belongs to the orthogonal subspace $\mathrm{span}\{H,\{X_j\}\}_{\omega}^\perp$.
Considering the analogy between the operators $\{H,\{X_j\},\{\bar P_k\}\}$ and basis vectors in a vector space, specifying $E$ fixes the direction $H$ in the space of operators, while specifying $\{x_j\}$ fixes the directions of all $X_j$.
The directions $\bar P_k$ are fixed by the $[d-(d_T+1)]$ independent conserved quantities other than $E$ and $\{x_j\}$. 

Any observable of the system can be written as $O=\sum_k o_k P_k+\sum_{l\neq k}o_{kl}|k\rangle\langle l|$,
but notably,
\[
\begin{aligned}
  \tr\{O \bar\rho\} &= \tr\{O_\mathrm{diag} \bar\rho\}
  \\[0.2cm]
  \tr\{O \omega\} &= \tr\{O_\mathrm{diag} \omega\},
\end{aligned}
\]
where $O_\mathrm{diag} = \sum_k o_k P_k$.
By the transformation leading to~\eqref{eq:maxentrhobar2}, we have~\cite{Sirker2014}
\[
 O_\mathrm{diag} = a_H(O)H + \sum_{j=1}^{d_T} a_{X_j}(O) X_j + \sum_{k=d_T+1}^d a_k(O)\bar{P}_k, 
\]
where
\[
\begin{aligned}
 a_{X_j}(O) = \frac{\tr\{O_\mathrm{diag} X_j \omega\}}{\tr\{X_j^2 \omega\}}, \quad
 a_k(O) = \frac{\tr\{ O_\mathrm{diag} \bar{P}_k \omega \}}{\tr\{\bar P_k^2 \omega\}},
\end{aligned}
\]
with $j=0,1,\dots,d_T$ and $X_0=H$.

Since $E = \tr\{H\bar\rho\} = \tr\{H\omega\}$ and $x_j = \tr\{X_j\bar\rho\} = \tr\{X_j \omega\}$, then, by construction,
\begin{equation}\label{eq:Odiffs}
  \tr\{O\bar \rho\} - \tr\{O \omega\} = \sum_{k=d_T+1}^d a_k(O) \tr\{\bar P_k (\bar\rho - \omega)\},   
\end{equation}
which explicitly depends on the directions $\bar{P}_k$ determined by conserved quantities other than $E$ and $\{x_j\}$.

The decomposition~\eqref{eq:Odiffs} is the basis for the definition of thermalization in Ref.~\cite{Sirker2014}, and is rooted in a distinction between local and nonlocal conserved charges.
A local observable is one that can be written in the thermodynamic limit as a sum/integral of operators with bounded spatial support.
When the local observable commutes with the Hamiltonian, it constitutes a local conserved charge.
Thus, in~\cite{Sirker2014}, a system is said to thermalize if, in the thermodynamic limit $N\to\infty$, for every local observable $O$, the expectation value in the dynamical equilibrium state $\bar\rho$ coincides with that in the corresponding Gibbs ensemble ($\omega$) constructed from the local conserved charges.

Here, this thermalization condition is equivalent to demanding that
\[
 \sum_{k=d_T+1}^d a_k(O)\frac{\tr\{\bar P_k (\bar\rho - \omega)\}}{N} \to 0\, \text{\,as\,}\, N\to\infty,
\]
for all local observables $O$.
In other words, thermalization occurs when the components of $\bar\rho-\omega$ along the directions $\bar P_k$ associated with conserved quantities other than $E$ and $\{x_j\}$ become irrelevant in the thermodynamic limit for all local observables.
In that case, the expectation values of these observables are reproduced by the ensemble $\omega$, completely characterized by $\{E,\{x_j\}\}$, even though the dynamical equilibrium state $\bar\rho$ may still retain information about additional conserved quantities.

\bibliography{references}

\begin{thebibliography}{43}
\bibitem [{not()}]{note2again}%
  \BibitemOpen
  \href@noop {} {}\bibinfo {howpublished} {Formally we should write, instead of Eqs.~\eqref{eq:ABHam} and~\eqref{eq:Xdecomp}, $H = H_A\otimes\mathds{1}_B + \mathds{1}_A\otimes H_B + H_I$ and $X_j = X_j^A\otimes\mathds{1}_B + \mathds{1}_A\otimes X_j^B$, respectively. But for simplicity of notation, we will omit the identities everywhere. Likewise we use $E_A=\tr\{H_A\,\rho\}$ and $x_j^A=\tr\{X_j^A\rho\}$ instead of the accurate $E_A = \tr\{H_A\otimes\mathds{1}_B\,\rho\}=\tr\{H_A\,\rho_A\}$ and $x_j^A=\tr\{X_j^A\otimes\mathds{1}_B\,\rho\} = \tr\{X_j^A\,\rho_A\}$, with $\rho_A=\tr_B\{\rho\}$, and similarly for $E_B$ and $x_j^B$.}\BibitemShut {Stop}%
\end{thebibliography}

\end{document}